%% file: manuscript.tex
\documentclass[acmtog]{acmart}

\usepackage{booktabs} % For formal tables

\usepackage{tabularx}
\newcolumntype{g}{>{\columncolor{G}}c}
\newcolumntype{x}{>{\centering\arraybackslash}X}

\usepackage{siunitx}  % SI units
\usepackage{subcaption}
\usepackage{mathbbol}

\usepackage{calc}
\usepackage{color,colortbl}
\definecolor{R}{cmyk}{0, 1, 1, 0}
\definecolor{B}{cmyk}{1, 0, 0, 0.25}
\definecolor{Y}{HTML}{FFFCB3}
\definecolor{N}{HTML}{B3E4BD}
\definecolor{P}{HTML}{FFB3B3}
\definecolor{M}{cmyk}{0, 1, 0, 0}
\definecolor{G}{gray}{0.5}

\newcommand{\myparagraph}[1]{\vspace{1em}\noindent\textbf{#1}}
\newcommand{\textover}[3][l]{\makebox[\widthof{#3}][#1]{#2}}
\newcommand{\re}[1]{#1}

\graphicspath{{figs/}}
\usepackage[ruled]{algorithm2e} % For algorithms

\SetAlFnt{\small}
\SetAlCapFnt{\small}
\SetAlCapNameFnt{\small}
\SetAlCapHSkip{0pt}
\IncMargin{-\parindent}

\acmJournal{TOG}
\acmYear{2017}
\acmVolume{36}
\acmNumber{4}
\acmArticle{1}
\acmMonth{7}

\setcopyright{acmlicensed}
\acmDOI{http://dx.doi.org/10.1145/3072959.3073633}

\begin{document}

\title{\re{Perceptual Evaluation of Liquid Simulation Methods}}

\author{Kiwon Um}
\orcid{0000-0002-4139-9308}
\affiliation{\institution{Technical University of Munich}}
\email{kiwon.um@tum.de}

\author{Xiangyu Hu}
\affiliation{\institution{Technical University of Munich}}
\email{xiangyu.hu@tum.de}

\author{Nils Thuerey}
\affiliation{\institution{Technical University of Munich}}
\email{nils.thuerey@tum.de}

\renewcommand\shortauthors{K. Um et al.}

\begin{abstract}
  This paper proposes a novel framework to evaluate fluid simulation methods
  based on crowd-sourced user studies in order to robustly gather large numbers
  of opinions. The key idea for a robust and reliable evaluation is to use a
  reference video from a carefully selected real-world setup in the user study.
  \re{By conducting a series of controlled user studies and comparing their
    evaluation results, we observe various factors that affect the perceptual
    evaluation. Our data show that the availability of a reference video makes
    the evaluation consistent.} We introduce this approach for computing scores
  of simulation methods as \emph{visual accuracy} metric. As an application of
  the proposed framework, a variety of popular simulation methods are evaluated.
\end{abstract}




% We no longer use \terms command
%\terms{Design, Algorithms, Performance}


\thanks{This work is supported by the \emph{ERC Starting Grant} 637014.

  Author's addresses: K. Um {and} N. Thuerey, Informatics 15, Technical
  University of Munich; X.Y. Hu, Chair of Aerodynamics and Fluid Mechanics,
  Technical University of Munich.}

\begin{teaserfigure}
  \includegraphics[width=\linewidth]{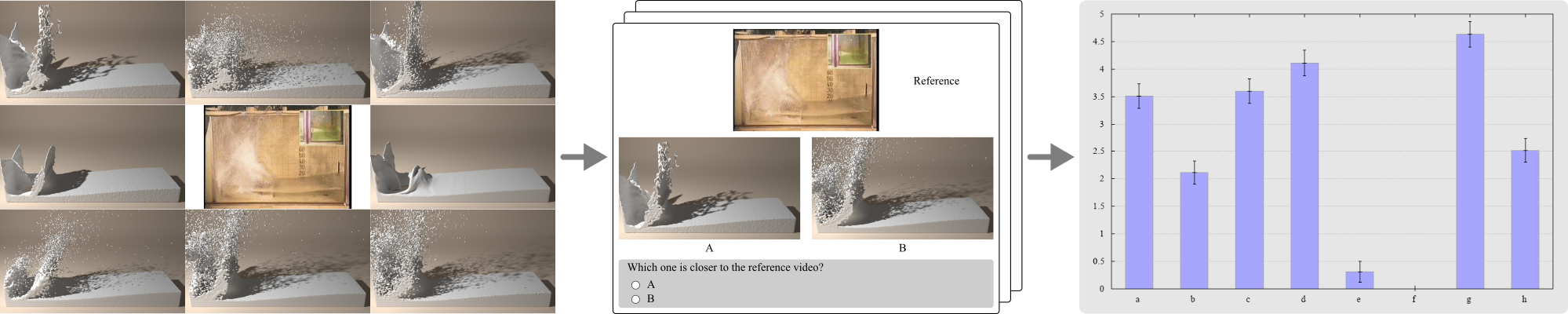}
  \caption{We evaluate different simulation methods (left) with user study
    consisting of pair-wise comparisons with reference (middle). This allows us
    to robustly evaluate the different simulation methods (right).}
  \label{teaser}
\end{teaserfigure}

\maketitle

\section{Introduction}
\label{sec:introduction}

In science, we constantly evaluate the results of our experiments. While some
aspects can be proven by mathematical measures such as the complexity class of
an algorithm, we resort to measurements for many practical purposes. When
measuring a simulation, the metrics for evaluation could be the computation time
of a novel optimization scheme or the order of accuracy of a new boundary
condition. These evaluation metrics are crucial for scientists to demonstrate
advances but also useful for users to select the most suitable one among various
methods for a given task.

This paper targets numerical simulations of liquids; in this area, most methods
strive to compute solutions to the established physical model, i.e., the
\emph{Navier-Stokes} (NS) equations, as accurately as possible. Thus,
researchers often focus on demonstrating an improved order of convergence to
show that a method leads to a more accurate solution
\cite{enright2003,kim2005,batty2007}. However, for computer graphics, the
overarching goal is typically to generate believable images from the
simulations. It is an open question how algorithmic improvements such as the
contribution of a certain computational component map to the opinion of viewers
seeing a video generated with this method.

There are several challenges here. Due to the complexity of our brain, we can be
sure that there is a very complex relationship between the output of a numerical
simulation and a human opinion. So far, there exist no computational models that can
approximate or model this opinion. \re{A second difficulty is that the transfer
  of information through our visual system is clearly influenced not only by the
  simulation itself but also by all factors that are involved with showing an
  image such as materials chosen for rendering and the monitor setup of a user.}
Despite these challenges, the goal of this paper is to arrive at a reliable
visual evaluation of fluid simulation methods. We will circumvent the former
problem by directly gathering data from viewers with user studies, and we will
design our user study setup to minimize the influence of image-level changes.

While there are interesting studies that investigate individual visual stimuli
\cite{han2016} and the influence of different rendering methods for liquid
simulations \cite{bojrab2013}, our goal is to calculate the perceptual scores
for fluid simulations on a high-level from animations produced with different
simulation methods. \re{We will demonstrate that a robust perceptual evaluation
  framework can be realized using crowd-sourced user studies that utilize
  carefully chosen simulation setups and a reference video.} This will allow us
to retrieve reliable \emph{visual accuracy} scores of different simulation
methods evaluated in each study. In order to establish this framework, we ran an
extensive series of user studies gathering more than 48,000 votes in total. The
overview of our framework is illustrated in Figure~\ref{teaser}.

In summary, we propose a novel perceptual evaluation framework for liquid
simulations. To the best of our knowledge, the perceptual evaluation of
physically-based liquid animations has previously not been studied, and we will
use our framework to evaluate different simulation methods and
parameterizations. From our evaluation results, we will draw useful observations
for different simulation methods.

\section{Related Work}

Fluid simulation methods typically compute solutions to the NS equations, which
can be written as
\re{$\partial\mathbf{u}/\partial{t} + \mathbf{u} \cdot \nabla \mathbf{u} =
  \mathbf{g} - \nabla{P}/\rho + \nu\nabla^2\mathbf{u}$ with the additional
  constraint to conserve volume: $\nabla\cdot\mathbf{u} = 0$, where $\mathbf{u}$
  is the velocity, $\mathbf{g}$ is the gravity, $P$ is the pressure, $\rho$ is
  the density, and $\nu$ is the viscosity coefficient.} Numerical solvers for
these equations can be roughly categorized as Eulerian and Lagrangian methods.
Fluid animations using Eulerian discretizations have been pioneered by Foster
and Metaxas~\shortcite{foster1996}, and the {\em stable fluids} solver
\cite{stam1999} has been widely used after its introduction. For liquids, the
particle level set method has been demonstrated to yield accurate and smooth
surface motions \cite{enright2002}. Currently, the fluid-implicit-particle
(FLIP) approach, which combines Eulerian incompressibility with a particle-based
advection scheme to represent small-scale details and splashes, is widely used
for visual effects \cite{zhu2005}. \re{The FLIP algorithm has been extended to
  many interesting applications such as the artistic control \cite{pan2013} and
  adaptivity \cite{ando2013}.} In the following, we will focus on liquid
simulations in simple domains without any adaptivity. We believe that this is a
good starting point for our studies, but these extensions would of course be
interesting for perceptual evaluations in the future.

The FLIP method was extended to incorporate position correction of the
participating particles \cite{ando2012,um2014} and to improve its efficiency by
restricting particles to a narrow band around the surface \cite{ferstl2016}.
\re{Secondary effects generation has been a highly popular topic within the
  fluid simulation area in order to increase the apparent detail of the
  simulation \cite{ihmsen2012}. Many movies and interactive applications have
  incorporated hand-tuned parameters and heuristics to approximate where and how
  splashes, foam, and bubbles develop from an under-resolved simulation.
  Moreover, a unilateral pressure solver was proposed to enable large-scale
  splashes in FLIP \cite{gerszewski2013}.} Recently, several more FLIP variants
were proposed to incorporate complex material effects that go beyond regular
Newtonian fluids \cite{stomakhin2013,ram2015}. We will later use the closely
related affine particle-in-cell (APIC) variant \cite{jiang2015} as one of our
candidates for simulation methods.

Lagrangian fluid simulation techniques in graphics are typically based on
variants of the smoothed particle hydrodynamics (SPH) approach. After its first
use for deformable objects \cite{debunne1999}, an SPH algorithm for liquids was
introduced by M\"uller et al. \shortcite{muller2003}, and then
weakly-compressible SPH (WCSPH) was introduced by Becker and Teschner
\shortcite{becker2007}. The SPH algorithm was adopted and extended in a
multitude of ways such as an adaptive discretization \cite{adams2007} and a
predictor-corrector step that improves efficiency and stability
\cite{solenthaler2009}. Techniques for two-way coupling between rigid bodies and
liquids have likewise been proposed \cite{akinci2012}.

A different formulation using the position-based dynamics viewpoint was proposed
for real-time simulations \cite{macklin2013} while other researchers suggested
an implicit method for better convergence rate \cite{ihmsen2014}; this is known
as implicit incompressible SPH (IISPH). From the Lagrangian field, we will
restrict our visual accuracy study to a few selected methods: WCSPH and IISPH,
which are typical and popular in graphics. Additionally, we also include an
engineering SPH variant \cite{adami2012}, from which we expect particularly
accurate simulations; we denote this variant as SPH in our studies.

Naturally, researchers have been interested in combining aspects of the
Lagrangian and Eulerian representations by bringing SPH and grid-based solving
components together \cite{losasso2008,raveendran2011}. We
have not yet included these hybrid approaches in our studies, although FLIP
arguably represents a hybrid particle-grid method. For a thorough overview of
popular fluid simulations methods, refer to the book by Bridson
\shortcite{bridson2015book} and state-of-the-art report by Ihmsen et al.
\shortcite{ihmsen2014star}.

The human visual system and perception of image and video contents have received
significant attention in computer graphics in order to study how algorithmic
choices influence the final judgment of the created images. For example, in the
area of rendering techniques, Cater et al. \shortcite{cater2002} proposed to use
selective and perceptually driven rendering approaches, and Dumont et al.
\shortcite{dumont2003} introduced a theoretical framework to compute perceptual
metrics. In photography, Masia et al. \shortcite{masia2009} perceptually
evaluated different techniques for tone-mapping HDR images with user studies.
For videos, an approach for perceptually-driven up-scaling of 3D content was
proposed \cite{didyk2010} while others investigated a
computational model for the perceptual evaluation of videos \cite{aydin2010}.

Beyond rendering and video, perceptual studies have also been used in the field
of character animation. Especially, human characters have received attention.
For instance, McDonnell et al. \shortcite{mcdonnell2008} studied how to
populate natural crowds for virtual environments. More recently, researchers
also gathered data on the attractiveness of virtual characters
\cite{hoyet2013}. In the area of deformable objects, Han and Keyser
\shortcite{han2016} studied how visual details can influence the perceived
stiffness of materials. Bojrab et al. \shortcite{bojrab2013} studied how
rendering styles of liquids influence user opinion. While this work also
considers liquids, our goal is in a way orthogonal to theirs. We focus on
simulation methods without being influenced by rendering styles.

\section{Visual Evaluation of Liquid Simulations}

Despite the fact that most liquid simulation methods are physically-based and
thus capable of approximating the NS equations in the limit, noticeable visual
differences exist among animations created from the different methods. Being
aware of these differences, we propose a novel approach that employs user
studies to evaluate the different methods in terms of how closely they match
real phenomena. The goal of our approach is to robustly and reliably compare
different liquid simulations such that the evaluation reflects a general
opinion. Therefore, we employ a crowd-sourcing platform in order to recruit many
participants to retrieve a reliable evaluation.

We focus on the perceptual evaluation of simulations in terms of what we call
\emph{visual accuracy}. We define this visual accuracy to be a score computed
from user study data to compare different methods, and we will make sure that it
can be computed in a robust and unbiased way. To collect data, we let users
select a preferred video from pair-wise comparisons, and we found it crucial for
robustness to provide participants with a visual reference. As we will outline
below, this also makes the results very stable with respect to strongly
differing rendering styles. These comparisons with a reference video are also
our motivation to see the scores we compute as a form of \emph{accuracy}.

Liquid simulations are commonly used tools in visual effects and applied for a
vast range of phenomena from drops of blood to large scale ocean scenes. While
it would be highly interesting to evaluate all of them, we focus on one
particular regime of water-like liquids on human scales. \re{This regime is
  highly challenging due to the low viscosity of water. The resulting flows
  typically feature high Reynolds numbers, complex waves, and large amounts of
  droplets and splashes. Although this naturally limits the regime of our study,
  we believe that it is particularly a representative for many effects and thus
  worth studying.} Next, we will present two carefully chosen simulation setups
that will also form the basis of our user studies in
Section~\ref{sec:user-study-method}.

\subsection{Simulation Setup}
\label{sec:sim-setup}

\begin{figure}[tb]
  \centering
  \begin{subfigure}[t]{0.48\linewidth}
    \centering
    \includegraphics[width=\linewidth]{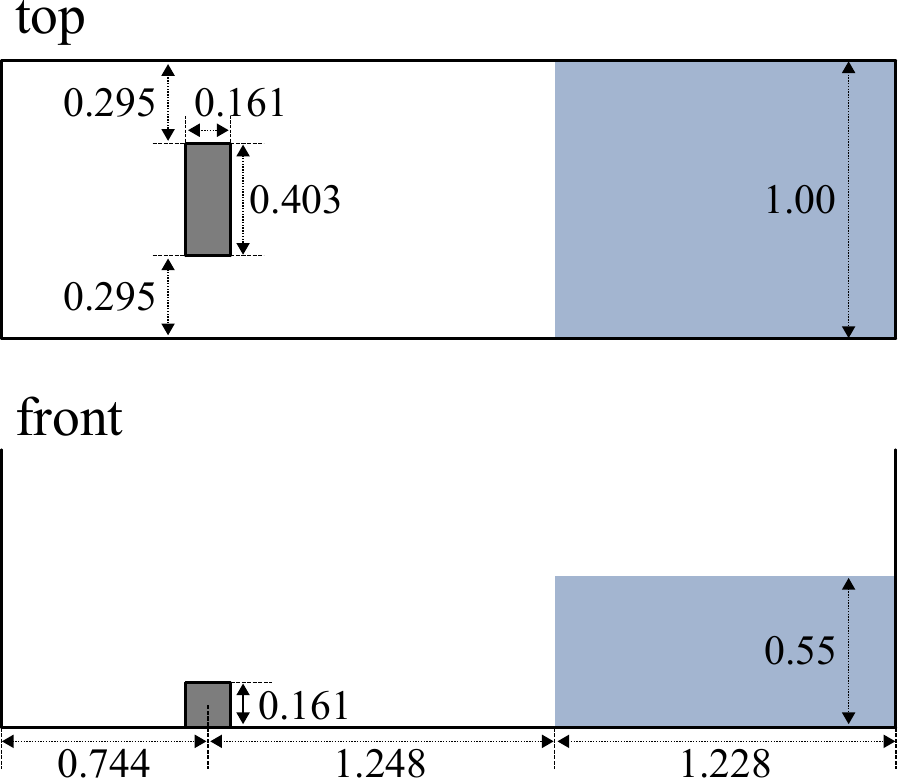}
    \includegraphics[width=\linewidth]{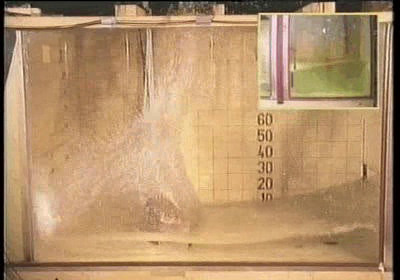}
    \caption{breaking-dam (in meter)}
  \end{subfigure}\hfill
  \begin{subfigure}[t]{0.48\linewidth}
    \centering
    \includegraphics[width=\linewidth]{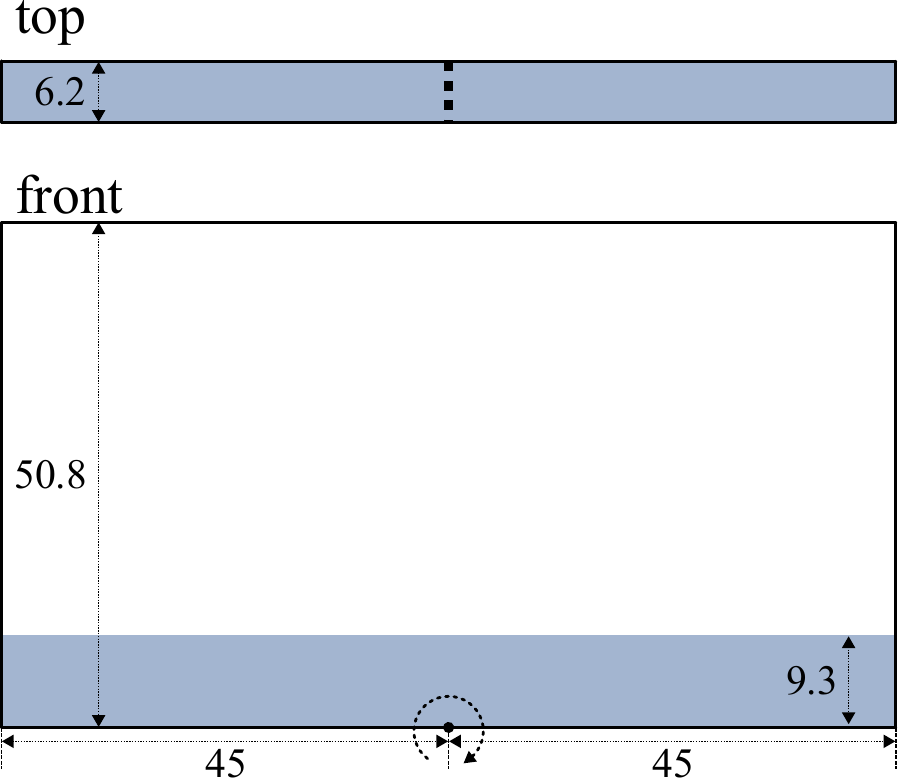}
    \includegraphics[width=\linewidth]{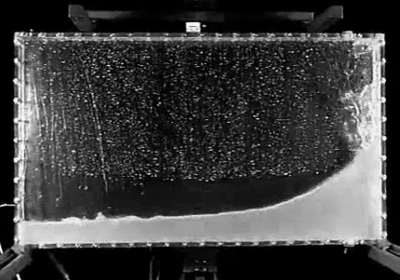}
    \caption{sloshing-wave (in centimeter)}
  \end{subfigure}
  \caption{Two simulation setups \re{\cite{kleefsman2005,botia2010}} for
    evaluation of liquid simulation and example frames of real experiments.}
  \label{fig:sim-setup}
\end{figure}

When selecting simulation setups, our requirements are that the setups are easy
to realize in numerical simulations; thus, they do not involve any specialized
domain boundary conditions or any moving obstacles. Therefore, the setups should
be easily reproducible. Nonetheless, the setups need to result in sufficiently
complex dynamics such as overturning waves and splashes in order to be relevant
for visual effects applications. Note that our setups stem from the engineering
community. This has the additional benefit that detailed flow measurements are
available as well as video data from real experiments. The latter is especially
important for our user study later.

Our first setup is close to the popular breaking dam case often seen in
graphics. Such a benchmark setup is also often used for validation in the
engineering studies, which adds an obstacle in front of the breaking dam for
additional complexity \cite{kleefsman2005}. This setup uses a tank of size
3.22\si{m}$\times$1\si{m}$\times$1\si{m} with an open roof, a static obstacle of
0.16\si{m}$\times$0.16\si{m}$\times$0.4\si{m}, and an initial water volume of
1.23\si{m}$\times$0.55\si{m}$\times$1.0\si{m}. As the tank is more than three
meters in length and the initial column is considerably high, this breaking dam
setup results in violent and turbulent splashes, which makes it tough but
relevant for our purposes. We will denote this setup as \emph{dam} in the
following, and the details of its initial conditions are illustrated in
Figure~\ref{fig:sim-setup}-(a).

Our second setup is a sloshing wave tank \cite{botia2010}; this is illustrated
in Figure~\ref{fig:sim-setup}-(b). A rectangular tank partially filled with
water experiences a periodic motion that continually injects energy into the
system leading to waves and splash effects forming over time. The size of the
tank is 0.9\si{m}$\times$0.51\si{m}$\times$0.062\si{m}, and the rotation axis is
located at the lower center of the tank. The initial water height is
0.093\si{m}. This setup has a significantly smaller overall water volume; it
leads to interesting waves forming over time. These waves are more prominent
here than in the dam setup. We will denote this setup as \emph{wave} in the
following. \re{Additional documentation for both setups is available online
  \cite{spheric}.}

For all simulations, we parameterize them according to the real-world dimensions
given above using earth gravity as the only external force. Unless otherwise
noted, we will not include any additional viscosity. In the following, we
explain the user studies, which are based on one of these two setups.

\subsection{User Study Design}
\label{sec:user-study-method}

\begin{figure}[tb]
  \centering
  \def\svgwidth{\linewidth}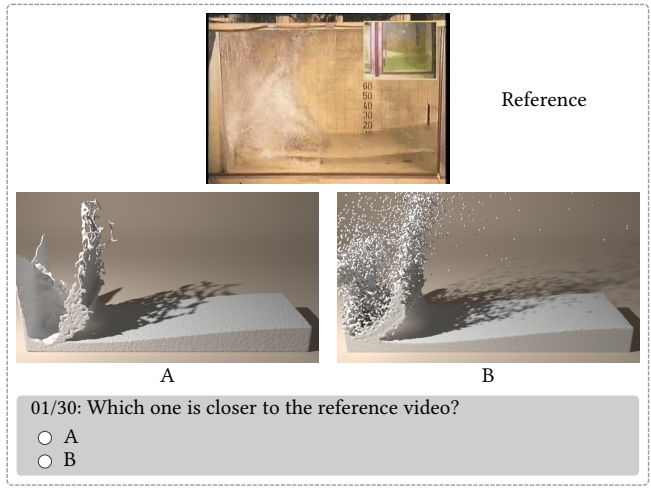
  \caption{Our user study design.}
  \label{fig:user-study}
\end{figure}

The goal of the user studies is to reliably evaluate the visual accuracy across
a set of $m$ videos produced by different simulation methods. While many
variants of user studies are imaginable \cite{leroy2011}, we opted for purely
binary questions in order to reduce noise and inconsistencies in the answers. We
also want to make the design as simple as possible to prevent misunderstandings.
Thus, participants are shown two videos to consider in comparison to a reference
video as illustrated in Figure~\ref{fig:user-study}. The videos are played
repeatedly without time limit, and the participants are given the task to select
one video which they consider to be closer to the reference video.

All participants have to give their vote for all possible pairs in a study.
Thus, for $m$ videos under consideration, we collect \mbox{$m(m-1)/2$} responses
per participant. In order to limit the workload per participant, we ensure that
$m$ is kept small, e.g., $m <= 7$ for our studies. In order to identify
untrustworthy participants, we duplicate the set of comparisons and randomize
their order; then, we check the consistency of the answers. We reject
participants with a consistency of less than 70\% \cite{cole2009}. Note that we
also randomize the positioning of both videos for each question (i.e., left and
right side).

Based on the pair-wise votes per study, we can now compute a set of scores for
all $m$ videos. For this purpose, we adopt the widely used Bradley-Terry model
\cite{bradley1952}. We review the model briefly here. Its goal is to compute
scores $s_i$ such that we can define the probability $p_{ij}$ that a participant
chooses video $i$ over video $j$ as:
\begin{equation}
  \label{eq:p-a-beat-b}
  p_{ij} = e^{s_i - s_j} / \left(1 + e^{s_i - s_j}\right).
\end{equation}
Let $w_{ij}$ denote the number of times where video $i$ was preferred over video
$j$ in a user study. Assuming the observations are independent, $w_{ij}$ follows
a binomial distribution. Therefore, the log likelihood for all pairs among all
videos can be calculated as follows:
\begin{equation}
  \label{eq:bt-likelihood}
  L(\mathbf{s}) = \sum_{i=1}^m\sum_{j=1}^m \left(w_{ij}s_i - w_{ij} \ln(e^{s_i} + e^{s_j})\right)
\end{equation}
where $\mathbf{s} = [s_1, s_2, ..., s_m]$. The final scores of all videos are
computed by solving for the $\mathbf{s}$ that maximizes the likelihood function
$L$ in Equation~(\ref{eq:bt-likelihood}) \cite{hunter2004}.

The vector of scores $\mathbf{s}$ is what we use to evaluate the visual accuracy
in the following. Note that these scores do not yield any ``absolute'' distances
to the reference, and they cannot be used to make comparisons across different
studies. \re{However, we found that they yield a reliable scoring and
  probability (see Equation~(\ref{eq:p-a-beat-b})) for all videos participating
  in a single study.}

In order to prevent bias with respect to the participants, we ran a series of
studies in three different crowd-sourcing platforms and found that differences
were negligible. Details for these studies can be found in
Appendix~\ref{sec:platforms}. Across our studies, we also noticed that the
consistency checks did not significantly influence the results, thus the large
majority of participants was trustworthy. In total, we collected user study data
for 48,800 pair-wise comparisons from 557 participants in 65 countries.

Seeing the consistency of answers across different platforms, we believe that
the user study design described above yields consistent answers. However, the
existence of consistent scores by themselves does not yet mean that we can draw
conclusions about the underlying simulation methods rather than about a certain
style of visualization. In the next section, we will present a series of user
studies to investigate whether we can specifically target simulation methods.

\subsection{Visual Accuracy for Simulations}
\label{sec:evaluations}

In order to show that there is a very high likelihood that our studies allow
conclusions to be drawn about the simulation methods, we now turn to comparisons
of studies. Thus, instead of considering individual visual accuracy scores
$s_i$, we will consider multiple sets of score vectors $\mathbf{s}$ to be
compared with each other. Once we have demonstrated that our user studies allow
us to draw conclusions with high confidence, we will discuss individual scores
for specific simulation-related questions in Section~\ref{sec:results}.

In the following, we will analyze pairs of studies for which we make only a
single change. For example, one study will have rendering style A, and a second
study will have rendering style B while keeping all other conditions identical.
We then perform a correlation analysis for these studies. If the studies turn
out to be correlated, we can draw conclusions about the influence of the change
on the outcome.

\begin{figure}[tb]
  \captionsetup[subfigure]{aboveskip=0pt,belowskip=0pt}
  \centering
  \begin{subfigure}[b]{0.49\linewidth}
    \centering
    \includegraphics[width=\linewidth]{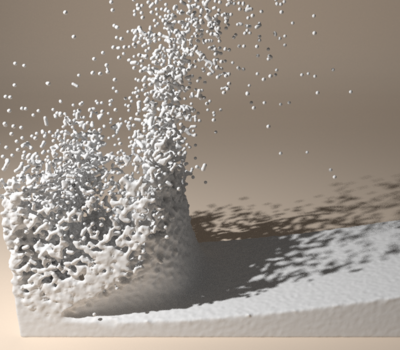}
    \caption{Opaque}
  \end{subfigure}
  \begin{subfigure}[b]{0.49\linewidth}
    \centering
    \includegraphics[width=\linewidth]{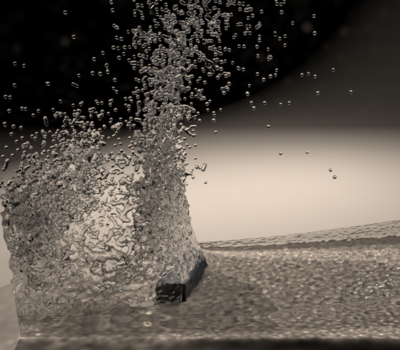}
    \caption{Transparent}
  \end{subfigure}
  \caption{Example frames of the opaque and transparent rendering styles.}
  \label{fig:opq-trp}
\end{figure}

For the correlation analysis, we compute the Pearson correlation coefficient and
statistical significance \cite{pearson1920}, which are widely used in statistics
as a measure of the linear correlation between two variables $x$, $y$
$\in \mathbb{R}^m$. This correlation coefficient $r$ is the covariance of the
two variables divided by the product of their standard deviations $\sigma_x$ and
$\sigma_y$, i.e., $r = \mathtt{cov}(x,y)/\sigma_x\sigma_y$. A strong positive
correlation, i.e., very similar score distributions, will result in values close
to $+1$, while uncorrelated or inverted scores, hence very different user
opinions, will result in correlations of $0$ or even negative correlations of
$-1$.

\begin{table*}[tb]
  \caption{Correlation analysis for the sets of scores evaluated from different user studies using FLIP and SPH. Here, ref. denotes the reference video.}
  \label{tab:Pearson-ref-role}
  \begin{minipage}[c]{0.67\linewidth}
    \vspace{0pt}
    \centering
    \begin{tabularx}{\linewidth}{c|c|l|xx}
      \toprule
      ID    & Comparison \re{(IDs in Table~\ref{tab:all-scores})} & \multicolumn{1}{c|}{Constant parameters}       & $r$                                         & p-value              \\
      \midrule
      C$_0$ & opaque (A) vs. transparent (B)                      & dam \textbf{\color{B}with ref.}                & \cellcolor{N}\textover[r]{0.97347}{$-$0.00000} & \cellcolor{N}0.00105 \\
      C$_1$ & dam (A) vs. wave (C)                                & rendered in opaque \textbf{\color{B}with ref.} & \cellcolor{N}\textover[r]{0.96557}{$-$0.00000} & \cellcolor{N}0.00176 \\
      \midrule
      C$_2$ & opaque (A*) vs. transparent (B*)                    & dam \textbf{\color{R}w/o ref.}                 & \cellcolor{Y}$-$0.01308                        & \cellcolor{Y}0.98039 \\ % both
      C$_3$ & dam (A*) vs. wave (C*)                              & rendered in opaque \textbf{\color{R}w/o ref.}  & \cellcolor{Y}\textover[r]{0.83895}{$-$0.00000} & \cellcolor{Y}0.03682 \\
      \midrule
      C$_4$ & with ref. (A) vs. w/o ref. (A*)                     & dam rendered in opaque                         & \cellcolor{P}\textover[r]{0.64540}{$-$0.00000} & \cellcolor{P}0.16632 \\
      C$_5$ & with ref. (B) vs. w/o ref. (B*)                     & dam rendered in transparent                    & \cellcolor{P}$-$0.60960                        & \cellcolor{P}0.19887 \\ % both
      \bottomrule
    \end{tabularx}
  \end{minipage}
  \begin{minipage}[c]{0.32\linewidth}
    \vspace{0pt}
    \centering
    \includegraphics[width=\linewidth]{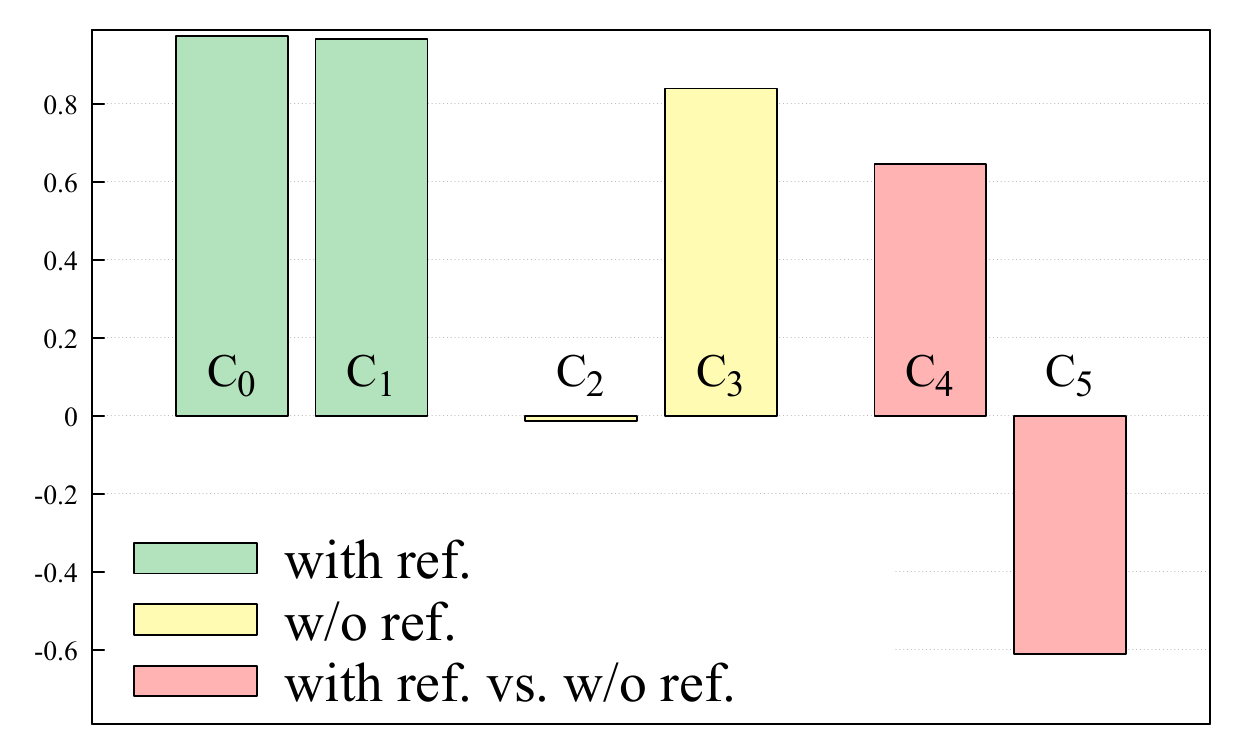}    
  \end{minipage}
\end{table*}

\begin{table}[b]
  \caption{Six simulation configurations for the experiments of Table~\ref{tab:Pearson-ref-role} and \ref{tab:Pearson-similar-visual}. Here, S denotes the scaling factors of resolution, and M denotes the methods: Eulerian (Eu.) and Lagrangian (La.)}
  \label{tab:dataset}
  \begin{tabularx}{\linewidth}{x|x|c|c}
    \toprule
    S           & M   & \multicolumn{2}{c}{Resolutions for particle and grid}                                                                                                                                                                               \\
    \cmidrule{3-4}
                &     & \emph{dam}                                                                                                        & \emph{wave}                                                                                                     \\
    \midrule
    1\textsf{x} & Eu. & \textover[r]{83k}{0,000k}    (\textover[r]{80}{000}$\times$\textover[r]{75}{000}$\times$\textover[r]{25}{000})    & \textover[r]{23k}{0,000k}    (\textover[r]{75}{000}$\times$\textover[r]{42}{000}$\times$\textover[r]{5}{00})    \\
    2\textsf{x} & Eu. & \textover[r]{664k}{0,000k}   (\textover[r]{160}{000}$\times$\textover[r]{150}{000}$\times$\textover[r]{50}{000})  & \textover[r]{186k}{0,000k}   (\textover[r]{150}{000}$\times$\textover[r]{84}{000}$\times$\textover[r]{10}{00})  \\
    4\textsf{x} & Eu. & \textover[r]{5,315k}{0,000k} (\textover[r]{320}{000}$\times$\textover[r]{300}{000}$\times$\textover[r]{100}{000}) & \textover[r]{1,488k}{0,000k} (\textover[r]{300}{000}$\times$\textover[r]{168}{000}$\times$\textover[r]{20}{00}) \\
    1\textsf{x} & La. & \textover[r]{84k}{0,000k}    (\textover[r]{80}{000}$\times$\textover[r]{75}{000}$\times$\textover[r]{25}{000})    & \textover[r]{24k}{0,000k}    (\textover[r]{75}{000}$\times$\textover[r]{42}{000}$\times$\textover[r]{5}{00})    \\
    2\textsf{x} & La. & \textover[r]{665k}{0,000k}   (\textover[r]{160}{000}$\times$\textover[r]{150}{000}$\times$\textover[r]{50}{000})  & \textover[r]{186k}{0,000k}   (\textover[r]{150}{000}$\times$\textover[r]{84}{000}$\times$\textover[r]{10}{00})  \\
    3\textsf{x} & La. & \textover[r]{2,253k}{0,000k} (\textover[r]{240}{000}$\times$\textover[r]{225}{000}$\times$\textover[r]{75}{000})  & \textover[r]{634k}{0,000k}   (\textover[r]{225}{000}$\times$\textover[r]{126}{000}$\times$\textover[r]{15}{00}) \\
    \bottomrule
  \end{tabularx}
\end{table}

In order to investigate the robustness of our visual accuracy evaluation, we set
up user studies with six videos. The six versions were chosen to broadly sample
the space of typical resolutions and simulation methods. For the studies of this
section, we are not particularly interested in the specific details of the
simulation methods as long as they are representative for commonly used methods
of graphics applications. \re{With this goal in mind, we will use a popular
  Eulerian method FLIP \cite{zhu2005} and Lagrangian method SPH \cite{adami2012}
  with three representative resolutions as shown in Table~\ref{tab:dataset}.
  Note that FLIP effectively is a hybrid Lagrangian-Eulerian method. However, we
  consider FLIP as Eulerian in our studies due to its Eulerian pressure solver,
  which is a key component of the algorithm.} We put an emphasis on visual
aspects with the studies described in the following section.

The space of possible visualization techniques for liquid animations is huge.
Many freely available renderers exist to create realistic images.
Real-time applications typically use specialized shaders for efficiency, and
visual effects in movies employ very refined compositions of many layers to
produce highly realistic visuals. Instead of trying to cover this whole space of
possibilities, we focus on two extremes of the spectrum: a fully opaque
rendering style and perfectly transparent surface. While the former employs a
simple diffuse material similar to a preview rendering, the transparent
rendering style exhibits complex lighting effects, such as refraction,
reflection, and caustics. A consequence is that the surface is very clearly
visible for the diffuse surface in contrast to the transparent rendering. Still
images for an example of these two rendering styles can be found in
Figure~\ref{fig:opq-trp}.

\myparagraph{Comparisons of user studies:} Assuming that our design for user
studies is reliable, we expect to see a strong correlation when comparing two
studies with these different rendering styles despite the differences in
appearance. This hypothesis is confirmed with a correlation coefficient of more
than $0.97$ with a high confidence level (p<0.01). The details for this
correlation calculation C$_{0}$ as well as the following ones can be found in
Table~\ref{tab:Pearson-ref-role}, and the full studies under consideration are
given in Table~\ref{tab:all-scores}. Considering the significantly different
images resulting from these two rendering styles, we believe that the strong
correlation is an encouraging result.

When removing the reference video from the user study design (C$_{2}$), i.e.,
only showing two videos of numerical simulations with the task to select the
``preferred'' version, the result changes drastically. Instead of a positive
correlation, we now see a nearly no correlation (i.e., $r=-0.013$). Thus,
without the availability of a reference video, the opinion between videos
changes very strongly when switching from the opaque to the transparent
rendering style.

While we used the \emph{dam} setup for the study above, we now repeat this
comparison keeping the rendering style constant (i.e., opaque) and comparing
simulation setups (\emph{dam} versus \emph{wave}). \re{When performing these
  studies with reference videos, we see a strong correlation of 0.97 (C$_{1}$)
  with a high confidence level (p<0.01), whereas the correlation slightly drops
  to 0.84 when the reference video is removed (C$_{3}$). The absence of a
  reference video does not necessarily lead to inconsistent results for all
  cases, rather there is an increased chance of ambiguity and substantially
  different responses.}

From the first two pairs of comparisons, we draw the conclusion that the
availability of a visual reference is crucial for a consistent evaluation of the
liquid motion. Having a reference video even stabilizes results from strongly
differing visualization styles as illustrated with \re{the studies of C$_{0}$}.
The reference video is also the reason why we believe our results do not
contradict previous work that found significant influence of rendering styles on
perception for animated water \cite{bojrab2013}. Regarding liquid motions in the
human-scale regime, our results indicate that the influence of rendering can be
made negligible by providing a visual reference. Note that our reference does
not need to closely match the rendering style used for the simulation videos.
The results are consistent even for significantly stylized and different
rendering styles such as our opaque and transparent styles; both are very
different from the reference video. The different correlation scores are
summarized visually at the figure in Table~\ref{tab:Pearson-ref-role}. This
figure again highlights that the low and even negative correlations are
stabilized by the availability of a reference video.

To shed further light on this topic, we compute correlations between the studies
with and without reference video. These correspondences can be found in
C$_{4,5}$ in Table~\ref{tab:Pearson-ref-role}. In both cases, the visual
accuracy scores of the methods under consideration change significantly when the
reference video is removed. \re{This results in the correlations that are not
  statistically significant (p>0.05). Besides, the results with the transparent
  rendering style show a drastic change of user opinions. Thus, without a
  reference video, visual appearance can strongly influence the scores.}

\begin{figure}[tb]
  \centering
  \includegraphics[width=\linewidth]{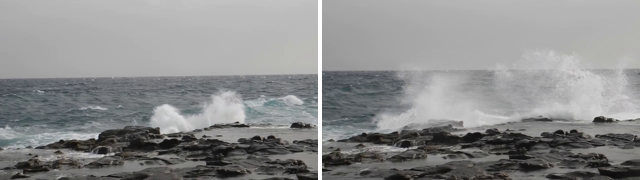}
  \caption{\re{Example frames of our alternate reference video for {\em dam}.}}
  \label{fig:seashore}
\end{figure}

\myparagraph{Reference videos:} \re{The video we used as reference for the
  \emph{dam} example has a visual appearance that is clearly different from our
  renderings. We note that visual accuracy can be evaluated even when the
  simulated phenomena bear only rough resemblance to the reference video.
  Figure~\ref{fig:seashore} shows the example frames of a reference video
  recorded in nature at a seashore. We use this video in an additional user
  study with the \emph{dam} example instead of the one shown in
  Figure~\ref{fig:sim-setup}, and the resulting scores are highly correlated
  with the results of the original study with the video of the \emph{dam}
  experiment. Here, the correlation is 0.93 with a high confidence level
  (p<0.01).
  
  On the other hand, when we use a video that differs more strongly, the user
  study results start to change. The correlation between a study using the
  \emph{wave} video with the \emph{dam} simulations and the original study is
  not statistically significant (p>0.05). To summarize, our results show that a
  reliable visual accuracy can be established even if no reference to the exact
  simulation setup is available. The human visual system is powerful enough to
  correlate the visual inputs despite different appearance. However, the
  stability of the results drops when the physics differ substantially.}

\begin{table*}[tb]
  \caption{Additional correlation analysis for two sets of simulation methods. Here, the \emph{dam} example is used with opaque style.}
  \label{tab:Pearson-similar-visual}
  \centering
  \begin{tabularx}{\linewidth}{l|c|l|xx}
    \toprule
    \multicolumn{1}{c|}{ID} & Comparison \re{(IDs in Table~\ref{tab:all-scores})} & \multicolumn{1}{c|}{Constant parameters} & $r$                  & p-value              \\
    \midrule
    C$_6$                   & FLIP\&SPH (A) vs. APIC\&IISPH (D)                   & with ref.                                & \cellcolor{N}0.96057 & \cellcolor{N}0.00230 \\
    C$_7$                   & FLIP\&SPH (A*) vs. APIC\&IISPH (D*)                 & w/o ref.                                 & \cellcolor{N}0.96932 & \cellcolor{N}0.00140 \\
    \midrule
    C$_8$                   & with ref. (D) vs. w/o ref. (D*)                     & APIC\&IISPH                              & \cellcolor{Y}0.72139 & \cellcolor{Y}0.10562 \\
    \bottomrule
  \end{tabularx}
\end{table*}

\myparagraph{Representative methods:} At this point, we also want to confirm our
assumption that the two initially chosen simulation methods are representative
for commonly used Eulerian and Lagrangian methods. We choose two different
methods from the Eulerian and Lagrangian classes: APIC \cite{jiang2015} and
IISPH \cite{ihmsen2014}. With these two methods, we performed new user studies
keeping the remainder of the user study and simulation setups constant; i.e.,
the simulations use the same resolutions of particle and grid as before
(Table~\ref{tab:dataset}). The strong positive correlation for this pair of
studies confirms our initial assumption (C$_6$ in
Table~\ref{tab:Pearson-similar-visual}). \re{Note that our two sets of
  simulation methods are also correlated in studies without a reference video
  (C$_7$). Presumably, this indicates that the participants' tendency in
  preference among the two classes of methods is fairly consistent. In this
  case, the individual scores of each method change substantially between the
  FLIP\&SPH and APIC\&IISPH sets. Thus, this makes it difficult to draw a
  conclusion among the different methods of each class. However, the correlation
  between the two sets of methods confirms our assumption that these methods
  cover the space of Eulerian and Lagrangian classes well.} In addition, we find
that the availability of a reference video affects the stability also in these
methods (C$_8$). \re{This is consistent with the aforementioned results
  indicating again that the absence of reference video results in a chance of
  ambiguity.}

\section{Applications and Results}
\label{sec:results}

In this section, we use our approach to evaluate the visual accuracy of various
simulation methods (Section~\ref{sec:methods7} and \ref{sec:similar-time}).
\re{We also demonstrate that our evaluation allows us to redeem heuristic
  approaches, such as the grid resolution for particle skinning
  (Section~\ref{sec:pskinning}), or algorithmic modifications, such as a splash
  model for FLIP simulations (Section~\ref{sec:splash}).}

\subsection{Liquid Simulation Methods}
\label{sec:methods7}

When establishing our evaluation framework, a central goal was to compare
simulation methods. In the following, we evaluate seven simulation methods from
the Eulerian and Lagrangian classes: marker-particles (MP) \cite{foster1996}, a
solver with level set surface tracking (LS) \cite{foster2001}, FLIP
\cite{zhu2005}, and APIC \cite{jiang2015} as representatives of Eulerian
methods; WCSPH \cite{becker2007}, IISPH \cite{ihmsen2014}, and a so-called
wall-boundary SPH method \cite{adami2012} as representatives of Lagrangian ones.
Note that this classification is primarily based on whether the method uses a
grid in the pressure solver. Using these seven methods, we simulate our two
simulation setups, i.e., \emph{dam} and \emph{wave} from
Section~\ref{sec:sim-setup}.

\begin{figure}[tb]
  \centering
  \includegraphics[width=\linewidth]{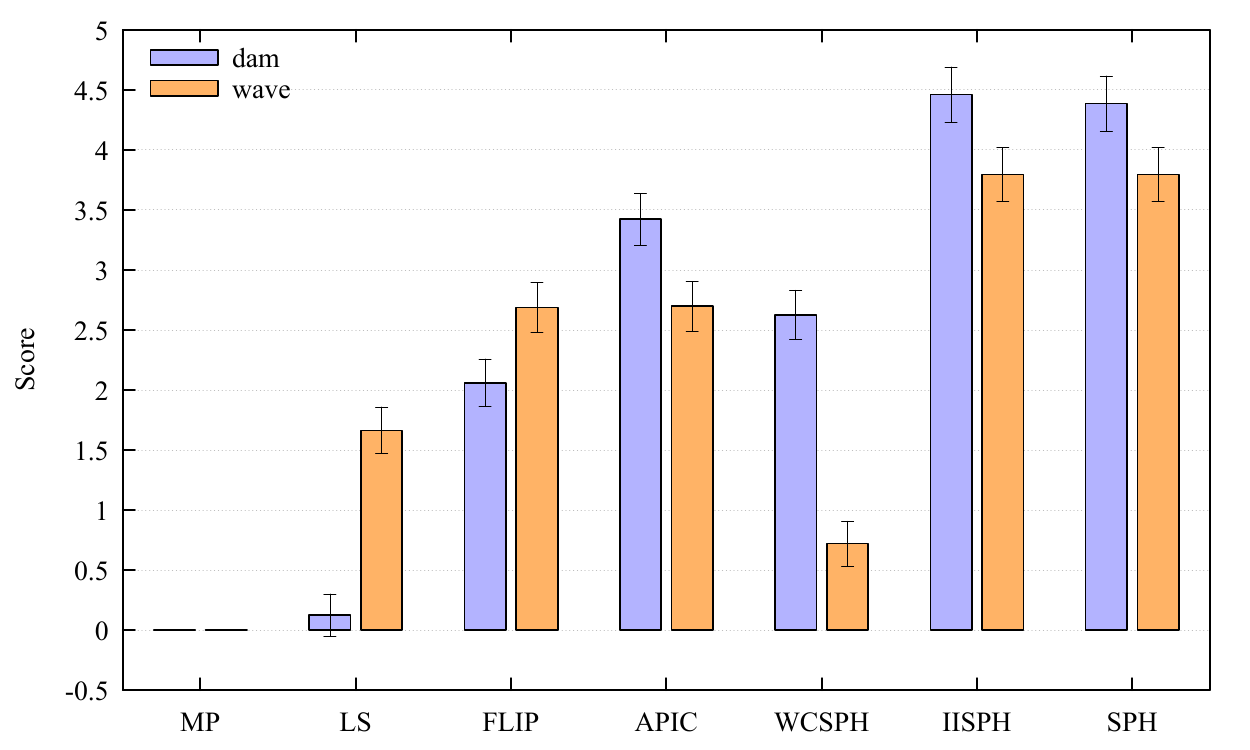}
  \caption{Visual accuracy scores of the seven simulation methods under
    consideration \re{(J and K in Table~\ref{tab:all-scores})}.}
  \label{fig:bt-methods}
\end{figure}

\begin{figure*}[tb]
  \centering
  \includegraphics[width=\linewidth]{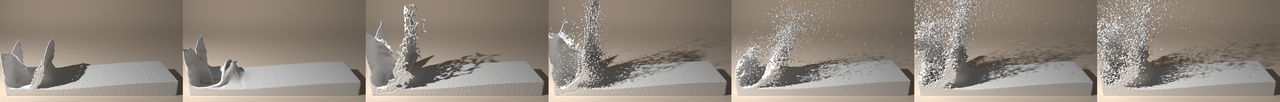}
  \includegraphics[width=\linewidth]{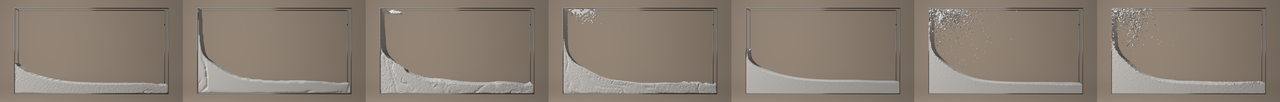}
  \caption{Example frames of seven simulations in two examples: (top) \emph{dam} and
    (bottom) \emph{wave}. From left to right, MP, LS, FLIP, APIC, WCSPH, IISPH and SPH
    are shown.}
  \label{fig:frames-methods}
\end{figure*}

The evaluation results are summarized in Figure~\ref{fig:bt-methods}.
Interestingly, the Lagrangian methods (particularly, IISPH and SPH) consistently
receive higher visual accuracy scores than the other methods. Among the Eulerian
methods, the FLIP variants (i.e., APIC and FLIP) receive higher scores than MP
and LS. Our guess for the latter results is that the MP and LS versions exhibit
a very small amount of droplets. Note that the score of WCSPH is also noticeably
low in the \emph{wave} example; we observe that the amount of splashes is
likewise very small, and the surface motion is highly viscous due to its
artificial viscosity. Here, the level set method receives a higher score than
WCSPH. We presume that this is caused by the artificial viscosity of the WCSPH
solve, which often results in a stronger damping of its surface motion in
comparison to the LS solve. Figure~\ref{fig:frames-methods} shows several still
frames of all the methods.

For the implementation of each method, we followed the original work without any
significant modifications. The Eulerian methods (i.e., MP, LS, FLIP, and PIC)
used grid resolutions of 160$\times$150$\times$50 for \emph{dam} and
150$\times$84$\times$10 for \emph{wave}. All methods except LS used 665K
particles for \emph{dam} and 186K particles for \emph{wave}. Although the
Lagrangian methods did not use any grid in their solve, we used the same
underlying grid for initializing the particles and sampled each cell with eight
particles. While the Lagrangian methods used a uniform sampling, the Eulerian
methods randomly jittered the particles to avoid aliasing. Note that this
resulted in slightly different numbers of particles ($\sim$1k) between the two
classes of methods; the particles were not reseeded during the simulation. In
order to ensure a comparable resolution for surface tracking in all methods, we
used a doubled resolution for tracking the level set in LS. For the pressure
solver of the Eulerian methods, we used a standard conjugate gradient method
with modified incomplete Cholesky preconditioning \cite{bridson2015book}.
\re{All implementations and setups can be found online.}

\subsection{Limited Computational Budget}
\label{sec:similar-time}

This experiment focuses on the four methods that ranked highest from the
previous evaluation and re-evaluates them with the constraint of a limited
computational budget per frame. While the previous study kept resolution and
particle count constant, we have adjusted them to yield comparable runtimes for
this study. We simulated the \emph{dam} example using APIC, FLIP, IISPH and SPH
such that they all required approximately 55 seconds per frame of animation.
\re{Here, we do not include the computational costs for non-simulation steps
  such as surface generation and rendering.} We are aware that absolute
comparisons of performance are difficult in general, but we have made our best
efforts to treat all methods fairly and to bring all implementations up to a
similar level of optimization (e.g., all implementations employ shared-memory
parallelism with OpenMP for most of their steps).

\begin{figure}[t]
  \captionsetup[subfigure]{aboveskip=0pt,belowskip=0pt}
  \centering
  \begin{subfigure}[b]{0.49\linewidth}
    \centering
    \includegraphics[width=\linewidth]{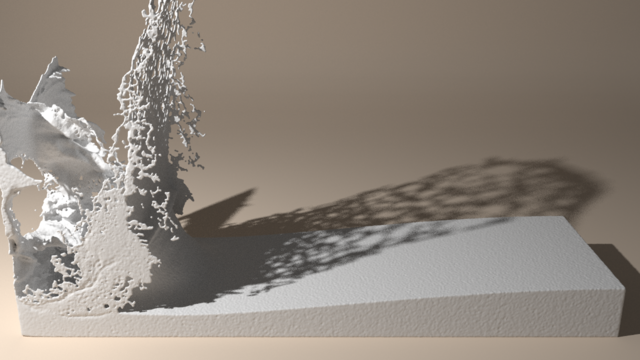}
    \caption{FLIP}
  \end{subfigure}
  \begin{subfigure}[b]{0.49\linewidth}
    \centering
    \includegraphics[width=\linewidth]{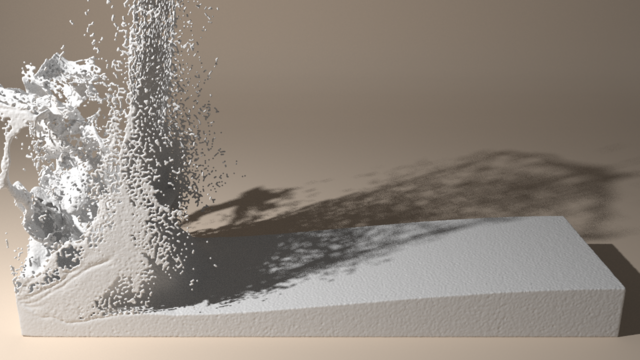}
    \caption{APIC}
  \end{subfigure}
  \begin{subfigure}[b]{0.49\linewidth}
    \centering
    \includegraphics[width=\linewidth]{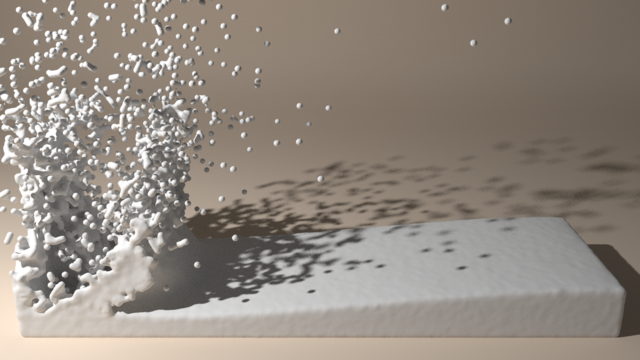}
    \caption{IISPH}
  \end{subfigure}
  \begin{subfigure}[b]{0.49\linewidth}
    \centering
    \includegraphics[width=\linewidth]{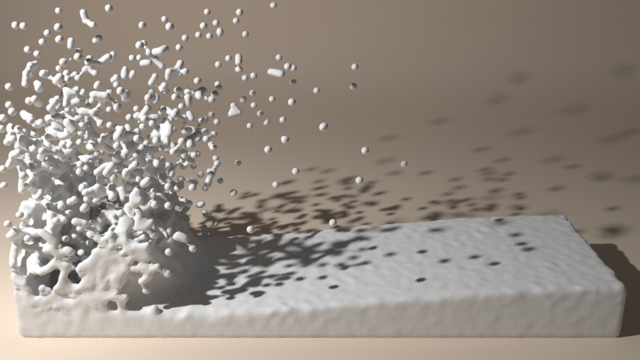}
    \caption{SPH}
  \end{subfigure}
  \caption{Example frames of four simulations with a similar computation time.}
  \label{fig:frames-time-limit}
\end{figure}

The time restriction leads to significant reduction in resolution for the
SPH-based methods. Both FLIP and APIC use a 320$\times$300$\times$100 grid and
5,315k particles; IISPH uses 143k particles sampled from a
96$\times$90$\times$30 grid, while SPH uses 84k particles sampled from a
80$\times$75$\times$25 grid. Example frames for these simulation configurations
are shown in Figure~\ref{fig:frames-time-limit}.

\begin{figure}[t]
  \centering
  \includegraphics[width=0.8\linewidth]{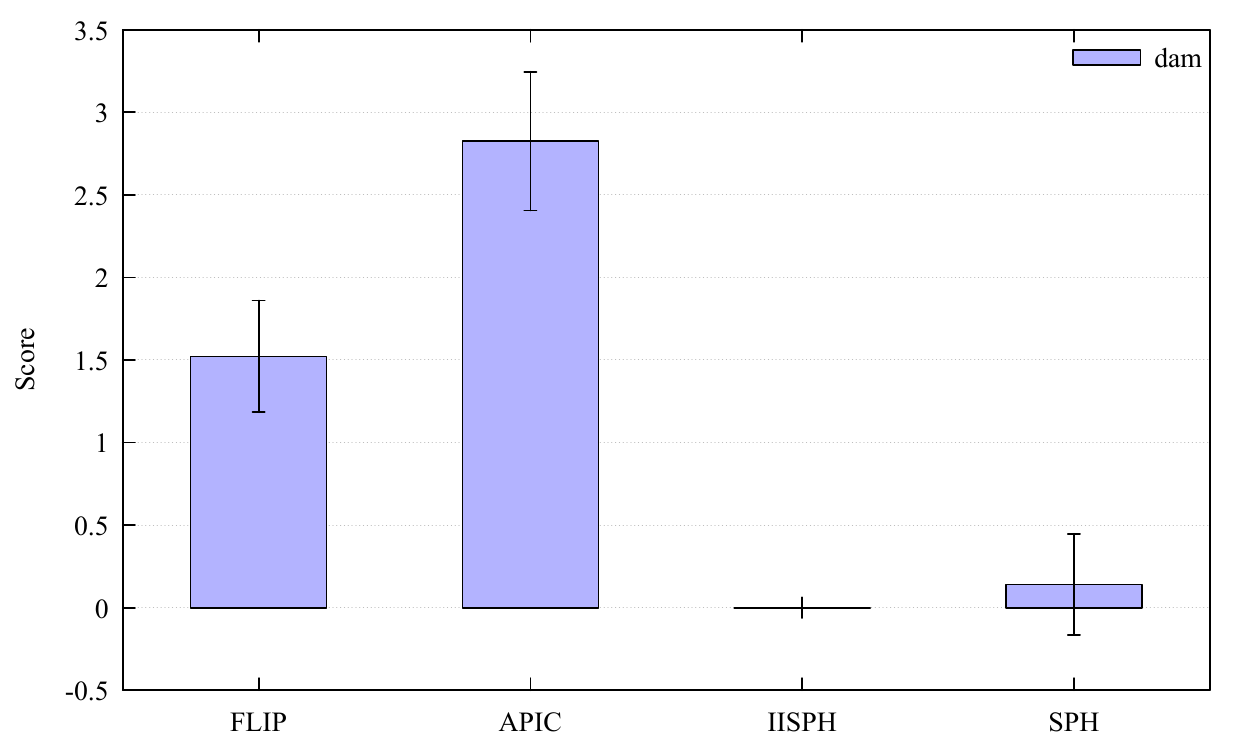}
  \caption{Visual accuracy scores of the four videos simulated in a similar
    computation time \re{(L in Table~\ref{tab:all-scores})}.}
  \label{fig:time-limit}
\end{figure}

In contrast to the previous evaluation in Section~\ref{sec:methods7}, our
participants gave the Eulerian methods higher visual accuracy scores. The
results are shown in Figure~\ref{fig:time-limit}. Thus, while the previous study
suggests that Lagrangian methods capture large-scale splashes better at a given
resolution, this study suggests that FLIP and APIC lead to improved results
under a restriction in computation time.

\subsection{Particle Skinning}
\label{sec:pskinning}

Our evaluation approach is also useful to redeem heuristic approaches, where
parameters are typically chosen by intuition. One example is the grid resolution
for generating a surface mesh from particle data, i.e., \emph{particle
  skinning}. The commonly used heuristic for this is to use a two times higher
resolution of the simulation grid, but there has been little motivation for this
particular setting.

\begin{figure*}[tb]
  \centering
  \includegraphics[width=\linewidth]{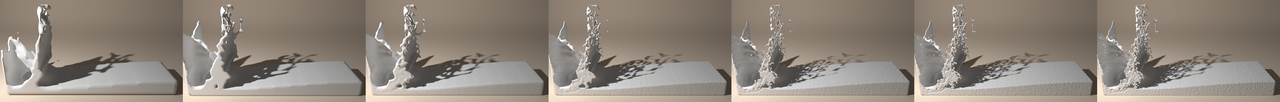}
  \caption{Example frames of seven resolutions for particle skinning. From left
    to right, 0.5\textsf{x}, 0.75\textsf{x}, 1\textsf{x}, 1.5\textsf{x},
    2\textsf{x}, 3\textsf{x}, and 4\textsf{x} are shown.}
  \label{fig:res-surf}
\end{figure*}

As the base simulation for this experiment, we use FLIP with a
160$\times$150$\times$50 grid and 664k particles. After simulation, a signed
distance field is computed from the particles \cite{zhu2005}, which we
triangulate with marching cubes. Since the particles are sampled at a 2$^3$
sub-grid, the cell size of the base resolution (1\textsf{x}) is 2$h$, where $h$
denotes the particle spacing. We perform the particle skinning using different
resolutions with seven scaling factors relative to $h$: 0.5\textsf{x},
0.75\textsf{x}, 1\textsf{x}, 1.5\textsf{x}, 2\textsf{x}, 3\textsf{x}, and
4\textsf{x}. In order to avoid missing particles in the grids that are more than
$h$ apart, the particle diameter is adjusted to the larger of either the grid
spacing or the particle spacing. The example frames are shown in
Figure~\ref{fig:res-surf}.

\begin{figure}[tb]
  \centering
  \includegraphics[width=0.8\linewidth]{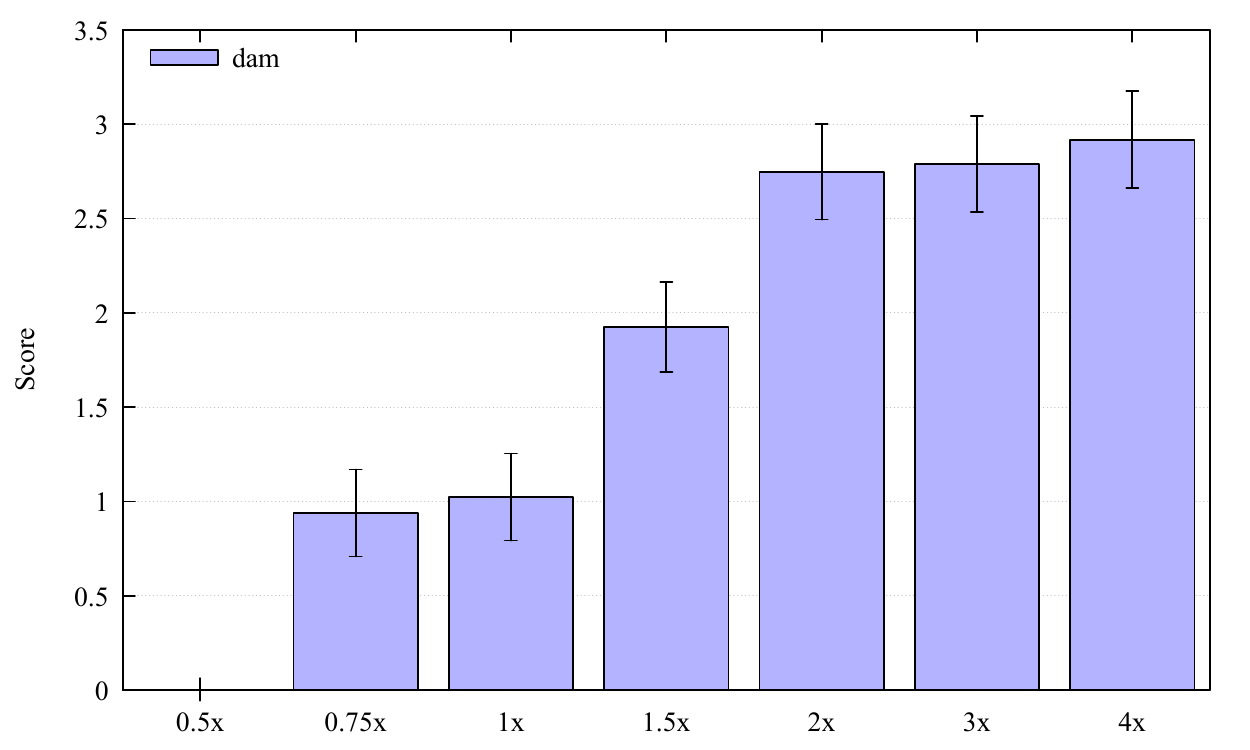}
  \caption{Visual accuracy scores of the seven resolutions for particle
    skinning \re{(M in Table~\ref{tab:all-scores})}.}
  \label{fig:bf-res-surf}
\end{figure}

As Figure~\ref{fig:bf-res-surf} shows, the evaluation result indicates that the
heuristic of 2\textsf{x} \cite{zhu2005} is a good one. The higher resolutions do
not yield results that can reliably be considered better than the 2\textsf{x}
factor, which thus represents the best performance.

\subsection{\re{Visual Impact of Splash Modeling}}
\label{sec:splash}

\begin{figure}[b]
  \captionsetup[subfigure]{aboveskip=0pt,belowskip=0pt}
  \centering
  \begin{subfigure}[b]{0.49\linewidth}
    \centering
    \includegraphics[width=\linewidth]{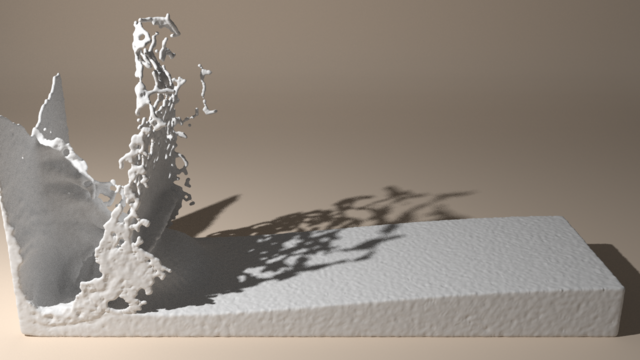}
    \includegraphics[width=\linewidth]{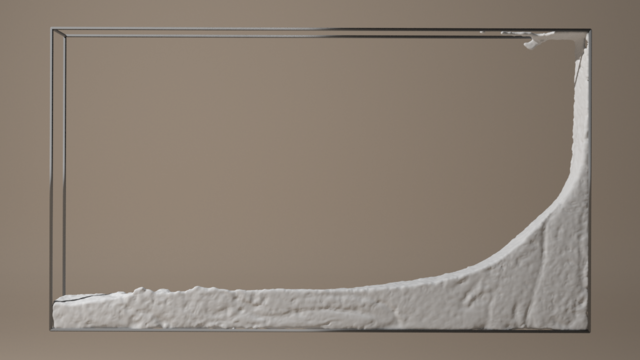}
    \caption{FLIP}
  \end{subfigure}
  \begin{subfigure}[b]{0.49\linewidth}
    \centering
    \includegraphics[width=\linewidth]{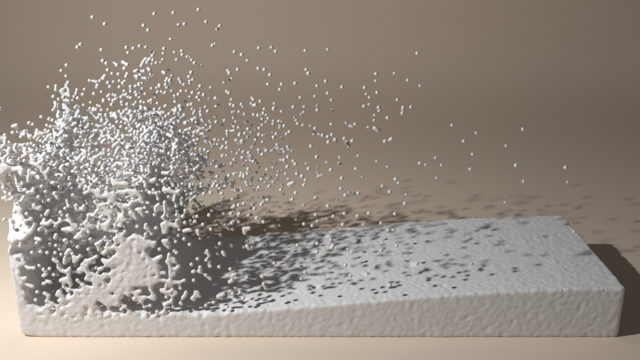}
    \includegraphics[width=\linewidth]{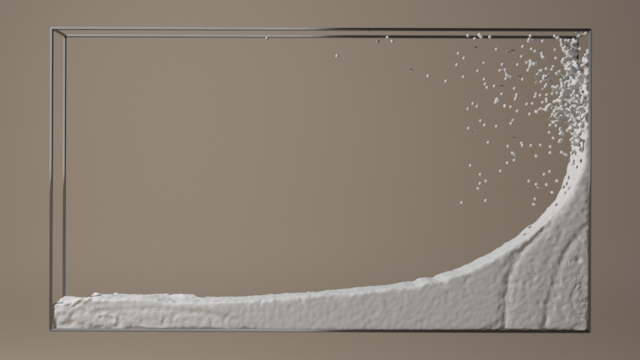}
    \caption{MLFLIP}
  \end{subfigure}
  \caption{Visual comparison of MLFLIP with FLIP in two examples: (top)
    \emph{dam} and (bottom) \emph{wave}.}
  \label{fig:frames-mlflip}
\end{figure}

\re{This section inspects a specific FLIP extension that claims to yield an
  increased amount of visual detail with secondary effects. It employs a
  neural-networks approach to model the sub-grid scale dynamics that lead to
  splashes \cite{um2017}, and we will denote it as \emph{MLFLIP} in the
  following. A visual comparison of example frames from both FLIP and MLFLIP can
  be seen in Figure~\ref{fig:frames-mlflip}.}

\begin{figure}[bh]
  \centering
  \includegraphics[width=\linewidth]{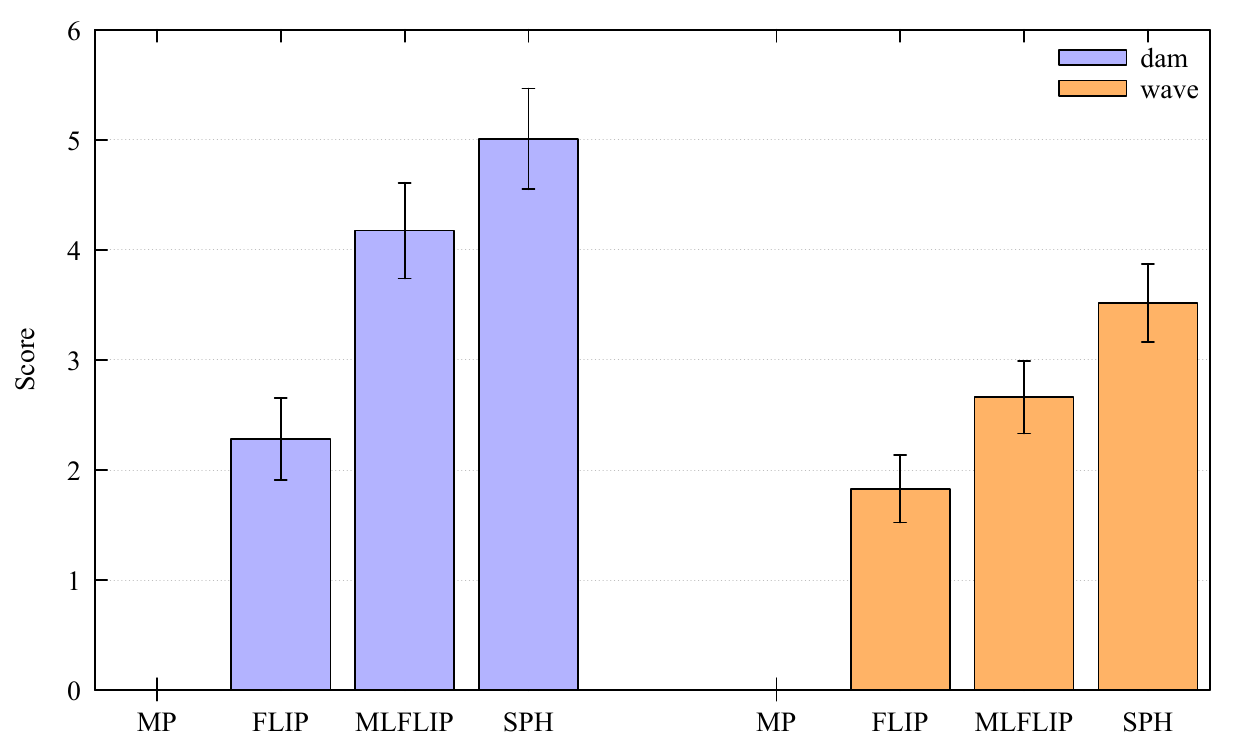}
  \caption{Notable improvements of MLFLIP in visual accuracy in two examples
    \re{(N and O in Table~\ref{tab:all-scores})}.}
  \label{fig:bt-methods-mlflip}
\end{figure}

In order to see whether this splash model indeed results in better visual
accuracy scores, we evaluate both FLIP and MLFLIP with two additional methods
for reference (i.e., MP and SPH). Figure~\ref{fig:bt-methods-mlflip} shows the
resulting visual accuracy scores. For the \emph{dam} setup, we observe that the
MLFLIP approach yields a notable improvement in score from 2.28 for regular FLIP
to 4.18 for MLFLIP. The gain for the \emph{wave} setup is lower, from 1.83 to
2.66, but we can still find a statistically relevant improvement. \re{These
  results indicate that splashes are an important visual cue for large-scale
  liquid phenomena.}

\section{\re{Discussion of Rendering Styles}}
\label{sec:renderings}

As our core method of evaluation, we propose to use measurements of visual
accuracy scores from user studies with a reference video. However, seeing the
strong variability in the previous results, especially for the transparent
rendering style, we believe that it is important to discuss additional studies
that we conducted to investigate the influence of rendering on the scores of
simulation methods when no reference video is available. However, we found this
area to be highly complex; thus, the following results are far from a complete
mapping of rendering space.

\begin{figure*}[tb]
  \captionsetup[subfigure]{aboveskip=0pt,belowskip=0pt}
  \centering
  \begin{subfigure}[b]{0.196\linewidth}
    \centering
    \includegraphics[width=\linewidth]{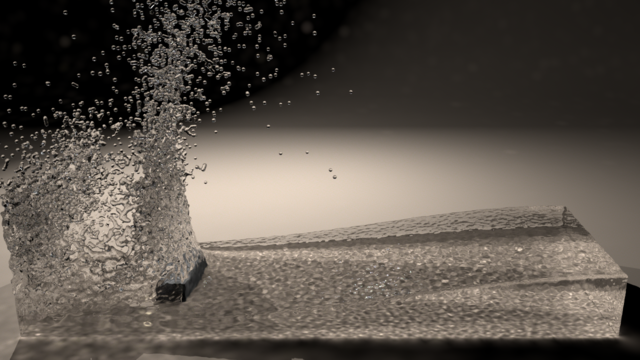}
    \caption{Transparent}
  \end{subfigure}
  \begin{subfigure}[b]{0.196\linewidth}
    \centering
    \includegraphics[width=\linewidth]{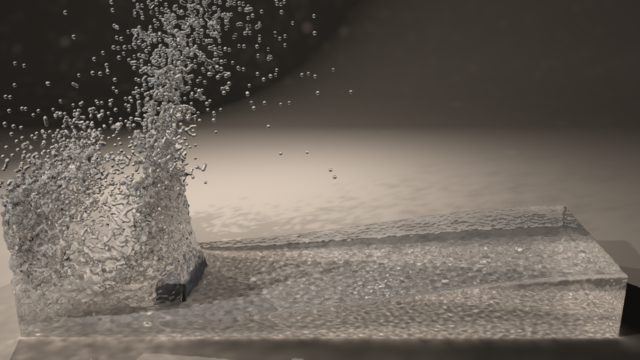}
    \caption{Blended (opaque: 0.25)}
  \end{subfigure}
  \begin{subfigure}[b]{0.196\linewidth}
    \centering
    \includegraphics[width=\linewidth]{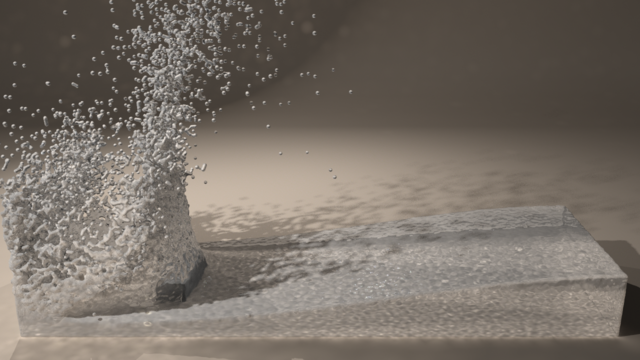}
    \caption{Blended (opaque: 0.5)}
  \end{subfigure}
  \begin{subfigure}[b]{0.196\linewidth}
    \centering
    \includegraphics[width=\linewidth]{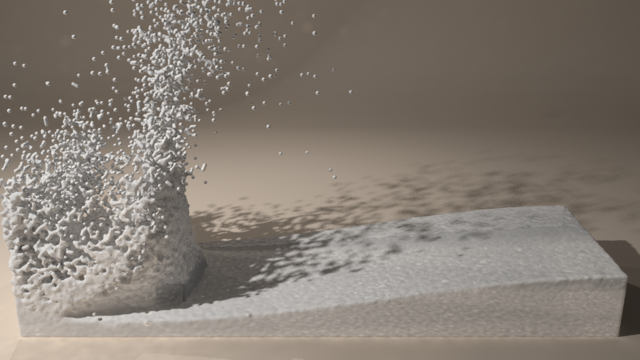}
    \caption{Blended (opaque: 0.75)}
  \end{subfigure}
  \begin{subfigure}[b]{0.196\linewidth}
    \centering
    \includegraphics[width=\linewidth]{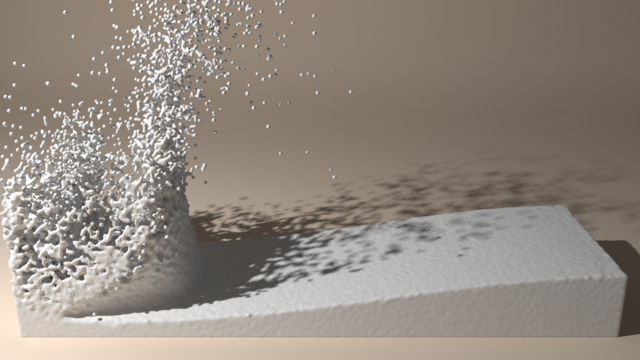}
    \caption{Opaque}
  \end{subfigure}
  \caption{Examples from our series of rendering styles transitioning from opaque to transparent.}
  \label{fig:rendering}
\end{figure*}

\begin{figure}[tbh]
  \centering
  \includegraphics[width=\linewidth]{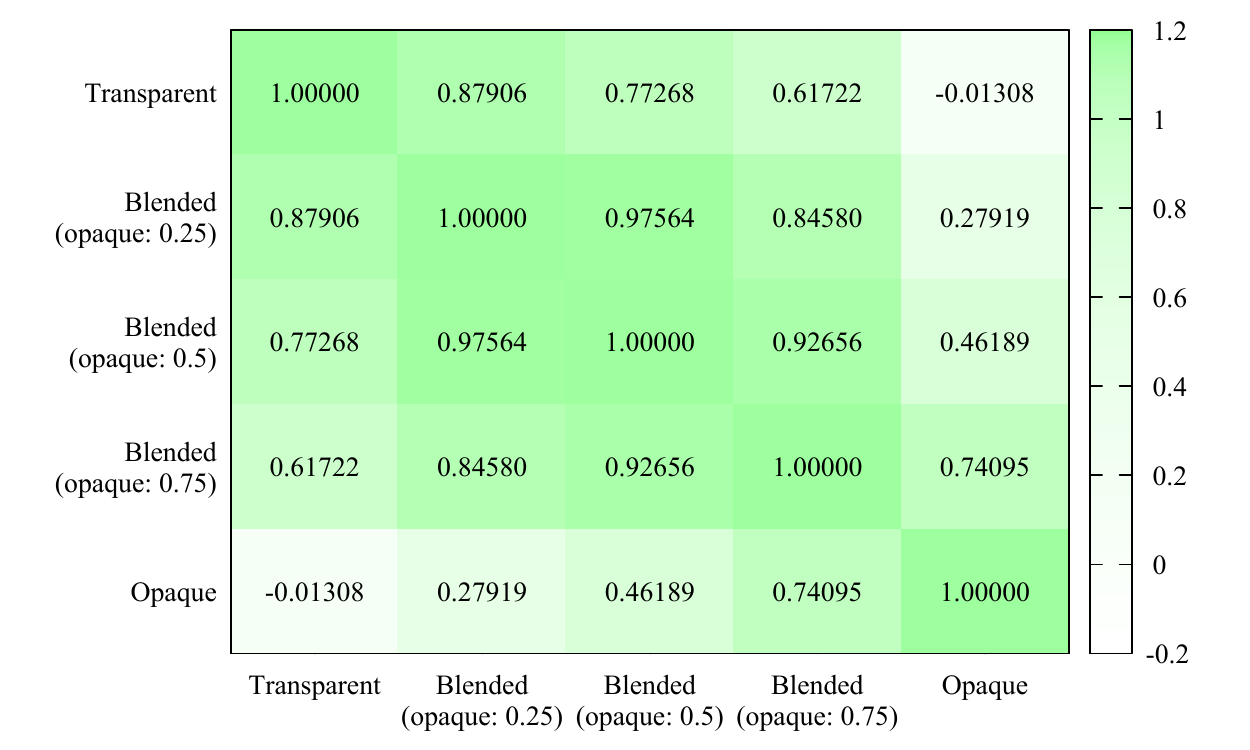}
  \caption{Correlation among the five sets of overall scores evaluated from the
    user studies with different rendering styles.}
  \label{fig:btscores-rendering}
\end{figure}

In a first series of studies, we investigate the behavior of the transition
between opaque and transparent rendering styles. We generated a sequence of
three in-between versions by linearly blending the two styles in image space as
shown in Figure~\ref{fig:rendering} and performed user studies. Interestingly,
the correlations between this series of studies change smoothly, albeit not
linearly, when moving from opaque towards transparent. The data are shown in
Figure~\ref{fig:btscores-rendering}. Due to the strong difference in initial
results (C$_2$ from Table~\ref{tab:Pearson-ref-role}), we found it surprising
that the space between these two extremes behaves smoothly.

\begin{figure}[b]
  \captionsetup[subfigure]{aboveskip=0pt,belowskip=0pt}
  \centering
  \begin{subfigure}[b]{0.49\linewidth}
    \centering
    \includegraphics[width=\linewidth]{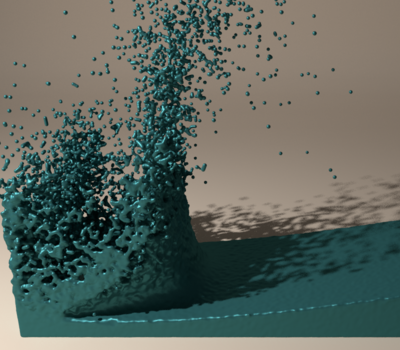}
    \caption{Glossy}
  \end{subfigure}
  \begin{subfigure}[b]{0.49\linewidth}
    \centering
    \includegraphics[width=\linewidth]{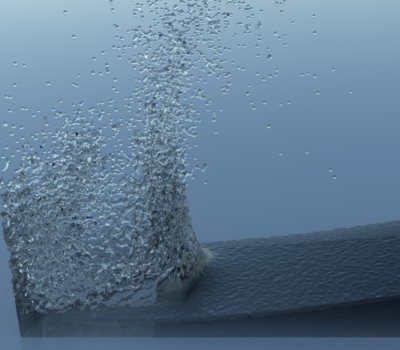}
    \caption{Translucent}
  \end{subfigure}
  \caption{Example frames of the rendering styles.}
  \label{fig:rendering-ext}
\end{figure}

We also performed the user studies with the same setup using two additional
rendering styles, which we selected to be different from both opaque and
transparent styles. The first additional style is a dark-green \emph{glossy}
surface, while the second one is a \emph{translucent} volume with attenuation
effects. These two rendering styles are shown in Figure~\ref{fig:rendering-ext}.
The correlation coefficients for these two styles with respect to our two
initial styles indicate that both the \emph{glossy} and \emph{translucent}
styles are strongly correlated with the opaque one as shown in
\re{Table~\ref{tab:rendering-ext}}. \re{Note that all studies discussed in this
  section were performed without the reference video. The results indicate that
  the opaque style covers a broader range of other rendering styles by showing
  the strong correlations even when no reference video is given.} Presumably,
the transparent rendering style with its complex light effects triggers a very
different ``mental image'' for the participants when no reference video is
given. This leads to a substantially different evaluation of the videos with
transparent rendering. \re{However, note that all studies in
  Section~\ref{sec:results} are conducted with the opaque rendering style and a
  reference since our goal is to reliably assess different methods.}

\begin{table}[tb]
  \caption{Correlation analysis for the additional rendering styles.}
  \label{tab:rendering-ext}
  \centering
  \begin{tabularx}{\linewidth}{c|xx}
    \toprule
    Comparison \re{(IDs in Table~\ref{tab:all-scores})} & $r$                  & p-value              \\
    \midrule
    Opaque (A*) vs. Glossy (H*)                         & \cellcolor{N}0.94329 & \cellcolor{N}0.00473 \\
    Opaque (A*) vs. Translucent (I*)                    & \cellcolor{N}0.93170 & \cellcolor{N}0.00684 \\
    Transparent (B*) vs. Glossy (H*)                    & \cellcolor{Y}0.55867 & \cellcolor{Y}0.24918 \\
    Transparent (B*) vs. Translucent (I*)               & \cellcolor{Y}0.59764 & \cellcolor{Y}0.21027 \\
    \bottomrule
  \end{tabularx}
\end{table}

\section{Conclusions and Outlook}

We have presented the first framework to perceptually evaluate liquid simulation
methods by employing crowd-sourced user studies. By analyzing the evaluation
results from controlled studies, we have demonstrated that our framework can
reliably measure user opinions in the form of a visual accuracy score. Our key
finding here is that the availability of a reference video makes stable
evaluations possible. Most importantly, the scores are not influenced by a
certain choice of rendering method.

The findings from our studies have led to several insights. For our chosen
settings, the studies suggest that
\begin{itemize}
\item viewers prefer SPH-based methods when comparable particle counts are used,
\item FLIP and especially APIC are preferred when the computational resources
  are limited,
\item the commonly used factor of two for particle skinning is confirmed by our
  experiment,
\item \re{and the splash effects are an important visual component for
    large-scale liquids.}
\end{itemize}

As the perception of physical phenomena such as liquids is highly complex, our
work clearly represents only a first step. We have not investigated the
demographics of our participants in more detail. Moreover, we currently focus on
a specific regime of liquid flows, and it is not clear how applicable our
results are for other regimes. Likewise, we have only tested a small selection
of simulation methods with our studies. There are many interesting variants that
could be evaluated in addition to our current selection. In the future, we are
also highly interested in extending our studies to smoke flows and other types
of materials such as objects undergoing elasto-plastic deformations. As we have
proposed a first perceptual evaluation framework for liquid simulation methods,
we believe these directions are very interesting avenues for future work.

\begin{acks}

  \re{We would like to thank all members of the graphics labs of TUM and IST
    Austria for the thorough discussions and the SPHERIC community for providing
    the experimental videos.}

\end{acks}

\vspace{2em}                    % solve the breaking bib entry problem

\bibliographystyle{ACM-Reference-Format}
\bibliography{references,references-perception}

\appendix

\section{Crowd-Sourcing Platforms}
\label{sec:platforms}

There exist several crowd sourcing services that provide a web-based platform where
the requester can launch user studies with a web-interface. This section
compares three popular platforms: \href{www.mturk.com}{Amazon Mechanical Turk}
(MT), \href{www.crowdflower.com}{CrowdFlower} (CF), and
\href{www.microworkers.com}{Microworkers} (MW).

In order to investigate consistency of all three platforms, we use our study
setup for \emph{dam} from Section~\ref{sec:evaluations} with six different
versions. In addition, we included an additional seventh dummy video, which was
synthesized by interleaving the six videos for each one second; we did not
include reverse questions in these three studies.

\begin{table}[bh]
  \caption{Three sets of scores evaluated from three platforms.}
  \label{tab:BT}
  \centering
  \begin{tabularx}{\linewidth}{c|xxx}
    \toprule
          & \multicolumn{3}{c}{Score (standard error)}          \\
    \cmidrule{2-4}
    ID    & CF              & MW              & MT              \\
    \midrule
    $a_1$ & 0.3317 (0.1637) & 0.6556 (0.1685) & 0.3677 (0.1661) \\
    $a_2$ & 0.2673 (0.1640) & 0.3539 (0.1693) & 0.0000 (0.1687) \\
    $a_3$ & 1.1146 (0.1665) & 0.9845 (0.1696) & 0.7923 (0.1666) \\
    $a_4$ & 1.6024 (0.1744) & 1.8701 (0.1820) & 1.8208 (0.1820) \\
    $a_5$ & 0.0000 (0.0000) & 0.0000 (0.0000) & 0.0000 (0.0000) \\
    $a_6$ & 0.8540 (0.1643) & 1.2118 (0.1715) & 1.0997 (0.1692) \\
    $a_7$ & 1.2931 (0.1688) & 1.7273 (0.1790) & 1.5849 (0.1766) \\
    \bottomrule
  \end{tabularx}
\end{table}

\begin{table}[bh]
  \caption{Pearson's correlations for the three platforms.}
  \label{tab:Pearson}
  \centering
  \begin{tabular*}{\linewidth}{@{\extracolsep{\fill}}cccc}
    \toprule
            & CF, MW  & MW, MT  & MT, CF  \\
    \midrule
    $r$     & 0.95808 & 0.98596 & 0.95170 \\
    p-value & 0.00068 & 0.00004 & 0.00096 \\
    \bottomrule
  \end{tabular*}
\end{table}

\begin{figure}[tb]
  \centering
  \includegraphics[width=\linewidth]{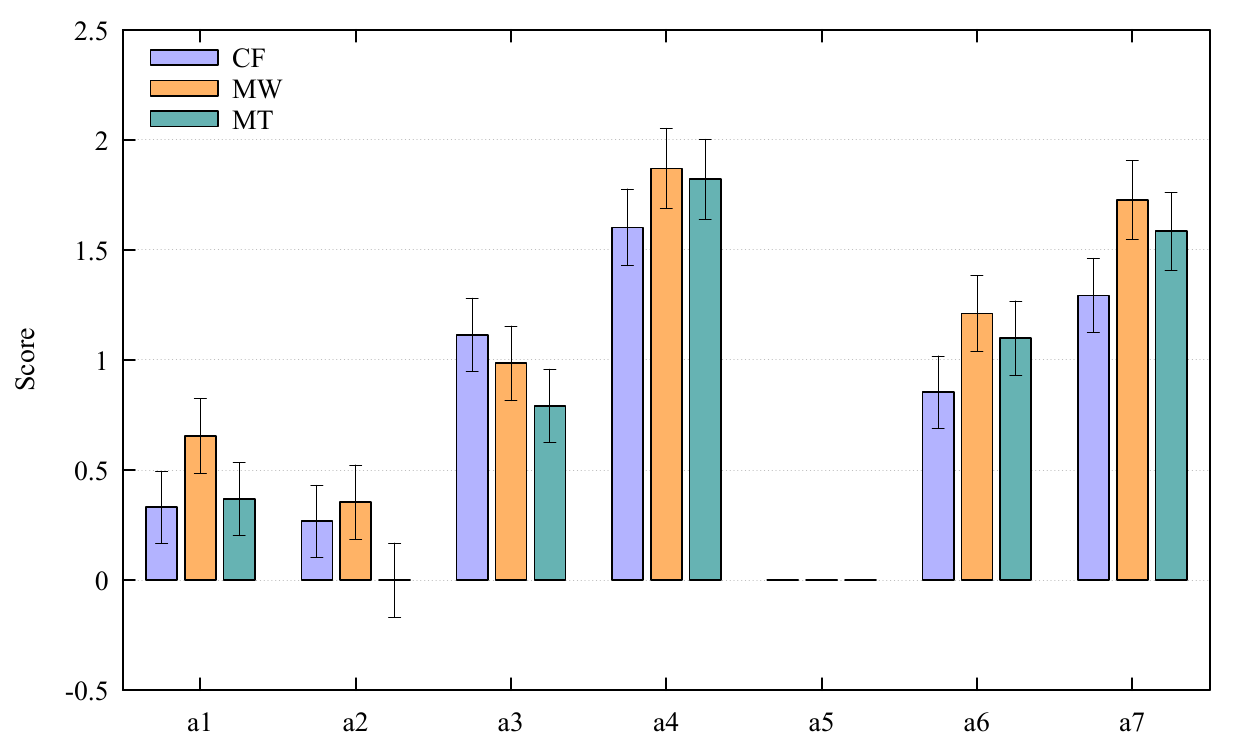}
  \caption{Graph of seven scores evaluated from three platforms.}
  \label{fig:btscores}
\end{figure}

Table~\ref{tab:BT} and Figure~\ref{fig:btscores} show the evaluation results
from the user study run on all three platforms, and Table~\ref{tab:Pearson}
shows the resulting correlation coefficients. As all p-values ($<$0.01) in the
table indicate, there is significant evidence with 99\% confidence to conclude
that the user studies obtained on the different platforms match. Thus, when only
considering the results of a single study, all three platforms yield very
similar results.

However, there are noticeable differences in the cost for each study. All
platforms allow the requester to set a cost for each query and the required
number of participants. An additional service fee is typically charged on top of
this. For our user study, we selected 50 participants for 21 queries and a per
query payment of \re{0.01 USD}, which resulted in costs of \re{21.00 USD} for
MT, \re{12.60 USD} for CF, and \re{23.10 USD} for MW. In addition, there were
significant differences in execution speed. With these settings, the MT platform
took several weeks to complete the study, while the other two platforms yielded
results in less than three days. Due to additional limitations with respect to
the maximal number of queries in the CF platform, we chose the MW platform for
all our studies.

\begin{table*}[tb]
  \caption{The visual accuracy scores (and standard errors).}
  \label{tab:all-scores}
  \centering
  \begin{tabularx}{\linewidth}{l|cx|rrrrrr}
    \multicolumn{9}{l}{Different examples and renderings (Section~\ref{sec:evaluations} and \ref{sec:renderings}), where A*, B*, D*, and E*-I* denote the studies without the reference video.} \\
    \toprule
    ID & Ex.  & Rendering   & FLIP, 1\textsf{x} & FLIP, 2\textsf{x} & FLIP, 4\textsf{x} & SPH, 1\textsf{x}   & SPH, 2\textsf{x}   & SPH, 3\textsf{x}                                            \\
    \midrule
    A  & dam  & opaque      & 0.0000 (0.0000)   & 3.1368 (0.4584)   & 4.6271 (0.4786)   & 4.9480 (0.4813)    & 6.5291 (0.4961)    & 6.7529 (0.4989)                                             \\
    A* & dam  & opaque      & 0.0000 (0.0000)   & 1.0822 (0.1975)   & 1.6328 (0.2083)   & 0.0579 (0.1964)    & 0.9089 (0.1955)    & 1.0300 (0.1968)                                             \\
    B  & dam  & transparent & 0.0000 (0.0000)   & 2.0498 (0.3272)   & 3.8288 (0.3572)   & 2.8715 (0.3428)    & 4.6016 (0.3700)    & 5.3260 (0.3864)                                             \\
    B* & dam  & transparent & 1.6860 (0.1785)   & 1.7125 (0.1789)   & 1.5685 (0.1765)   & 0.8198 (0.1694)    & 0.4223 (0.1695)    & 0.0000 (0.0000)                                             \\
    C  & wave & opaque      & 0.0000 (0.0000)   & 3.2189 (0.4720)   & 3.6823 (0.4771)   & 3.0738 (0.4701)    & 5.2235 (0.4996)    & 5.0324 (0.4958)                                             \\
    \midrule
    \midrule
       &      &             & APIC, 1\textsf{x} & APIC, 2\textsf{x} & APIC, 4\textsf{x} & IISPH, 1\textsf{x} & IISPH, 2\textsf{x} & IISPH, 3\textsf{x}                                          \\
    \cmidrule{4-9}
    D  & dam  & opaque      & 0.0000 (0.0000)   & 2.6095 (0.3411)   & 3.7208 (0.3541)   & 2.6466 (0.3416)    & 4.2966 (0.3618)    & 4.9892 (0.3751)                                             \\
    D* & dam  & opaque      & 0.1480 (0.1816)   & 1.5857 (0.1857)   & 2.0321 (0.1933)   & 0.0000 (0.0000)    & 1.4117 (0.1835)    & 1.8044 (0.1890)                                             \\
    \midrule
    \midrule
       &      &             & FLIP, 1\textsf{x} & FLIP, 2\textsf{x} & FLIP, 4\textsf{x} & SPH, 1\textsf{x}   & SPH, 2\textsf{x}   & SPH, 3\textsf{x}                                            \\
    \cmidrule{4-9}
    E* & dam  & blend 25\%  & 0.0000 (0.0000)   & -0.0776 (0.1762)  & 0.0000 (0.1769)   & -1.9924 (0.1972)   & -1.4552 (0.1849)   & -1.6837 (0.1895)                                            \\
    F* & dam  & blend 50\%  & 0.0000 (0.0000)   & 0.1456 (0.1629)   & 0.2132 (0.1636)   & -1.2302 (0.1726)   & -0.6418 (0.1629)   & -0.7460 (0.1641)                                            \\
    G* & dam  & blend 75\%  & 0.0000 (0.0000)   & 0.8031 (0.1843)   & 1.1089 (0.1919)   & -1.0177 (0.1914)   & -0.2983 (0.1779)   & -0.2034 (0.1772)                                            \\
    H* & dam  & glossy      & 0.0000 (0.0000)   & 0.5613 (0.1489)   & 0.8232 (0.1524)   & -0.7548 (0.1553)   & 0.1286 (0.1465)    & 0.0537 (0.1465)                                             \\
    I* & dam  & translucent & 0.0000 (0.0000)   & 0.8324 (0.1484)   & 0.8324 (0.1484)   & -0.2321 (0.1456)   & 0.1135 (0.1437)    & 0.0723 (0.1438)                                             \\
    \bottomrule
  \end{tabularx}\vspace{1em}

  \begin{tabularx}{\linewidth}{l|c|xxxxxxx}
    \multicolumn{9}{l}{Seven simulation methods (Section~\ref{sec:methods7})}                                                               \\
    \toprule
    ID & Ex.  & MP              & LS              & FLIP            & APIC            & WCSPH           & IISPH           & SPH             \\
    \midrule                                                                                                              
    J  & dam  & 0.0000 (0.0000) & 0.1248 (0.1769) & 2.0613 (0.1962) & 3.4211 (0.2136) & 2.6271 (0.2039) & 4.4595 (0.2294) & 4.3855 (0.2280) \\
    K  & wave & 0.0000 (0.0000) & 1.6646 (0.1943) & 2.6871 (0.2058) & 2.6987 (0.2060) & 0.7209 (0.1876) & 3.7943 (0.2229) & 3.7943 (0.2229) \\
    \bottomrule
  \end{tabularx}\vspace{1em}

  \begin{tabularx}{\linewidth}{l|xxxx}
    \multicolumn{5}{l}{Four simulation methods for \emph{dam} in similar computation time (Section~\ref{sec:similar-time})} \\
    \toprule
    ID & FLIP            & APIC            & IISPH           & SPH                                                  \\
    \midrule
    L  & 1.5215 (0.3387) & 2.8256 (0.4205) & 0.0000 (0.0000) & 0.1410 (0.3070)                                      \\
    \bottomrule
  \end{tabularx}\vspace{1em}

  \begin{tabularx}{\linewidth}{l|xxxxxxx}
    \multicolumn{8}{l}{Seven grid resolutions for particle skinning (Section~\ref{sec:pskinning})}                                    \\
    \toprule
    ID & 0.5\textsf{x}   & 0.75\textsf{x}  & 1\textsf{x}     & 1.5\textsf{x}   & 2\textsf{x}     & 3\textsf{x}     & 4\textsf{x}     \\
    \midrule
    M  & 0.0000 (0.0000) & 0.9397 (0.2308) & 1.0235 (0.2310) & 1.9248 (0.2393) & 2.7473 (0.2533) & 2.7891 (0.2542) & 2.9170 (0.2572) \\
    \bottomrule
  \end{tabularx}\vspace{1em}

  \begin{tabularx}{\linewidth}{l|xxxx}
    \multicolumn{5}{l}{Four methods including MLFLIP (Section~\ref{sec:splash})} \\
    \toprule
    ID & MP              & FLIP            & MLFLIP          & SPH               \\
    \midrule
    N  & 0.0000 (0.0000) & 2.2833 (0.3723) & 4.1758 (0.4353) & 5.0077 (0.4569)   \\
    O  & 0.0000 (0.0000) & 1.8312 (0.3069) & 2.6612 (0.3282) & 3.5203 (0.3538)   \\
    \bottomrule
  \end{tabularx}
  
\end{table*}

\end{document}

%% file: figs/study-ref.pdf_tex
%% Creator: Inkscape inkscape 0.91, www.inkscape.org
%% Accompanies image file 'study-ref.pdf' (pdf, eps, ps)
\begingroup%
  \makeatletter%
  \providecommand\color[2][]{%
    \errmessage{(Inkscape) Color is used for the text in Inkscape, but the package 'color.sty' is not loaded}%
    \renewcommand\color[2][]{}%
  }%
  \providecommand\transparent[1]{%
    \errmessage{(Inkscape) Transparency is used (non-zero) for the text in Inkscape, but the package 'transparent.sty' is not loaded}%
    \renewcommand\transparent[1]{}%
  }%
  \providecommand\rotatebox[2]{#2}%
  \ifx\svgwidth\undefined%
    \setlength{\unitlength}{545.2bp}%
    \ifx\svgscale\undefined%
      \relax%
    \else%
      \setlength{\unitlength}{\unitlength * \real{\svgscale}}%
    \fi%
  \else%
    \setlength{\unitlength}{\svgwidth}%
  \fi%
  \global\let\svgwidth\undefined%
  \global\let\svgscale\undefined%
  \makeatother%
  \begin{picture}(1,0.75055026)%
    \put(0,0){\includegraphics[width=\unitlength,page=1]{study-ref.pdf}}%
    \put(0.25056459,0.16101889){\color[rgb]{0,0,0}\makebox(0,0)[b]{\smash{{\small{A}}}}}%
    \put(0.74946407,0.16101889){\color[rgb]{0,0,0}\makebox(0,0)[b]{\smash{{\small{B}}}}}%
    \put(0.0891416,0.06713131){\color[rgb]{0,0,0}\makebox(0,0)[lb]{\smash{{\small{A}}}}}%
    \put(0.0891416,0.03044752){\color[rgb]{0,0,0}\makebox(0,0)[lb]{\smash{{\small{B}}}}}%
    \put(0,0){\includegraphics[width=\unitlength,page=2]{study-ref.pdf}}%
    \put(0.0377843,0.11115185){\color[rgb]{0,0,0}\makebox(0,0)[lb]{\smash{{\small{01/30: Which one is closer to the reference video?}}}}}%
    \put(0.83681734,0.59163064){\color[rgb]{0,0,0}\makebox(0,0)[b]{\smash{{\small{Reference}}}}}%
    \put(0,0){\includegraphics[width=\unitlength,page=3]{study-ref.pdf}}%
  \end{picture}%
\endgroup%